\documentclass[aps, amsfonts, nofootinbib, notitlepage]{revtex4}
\usepackage{amsmath}
\usepackage{epsfig}
\usepackage{xcolor}
\usepackage{hyperref}
\usepackage{float}
\usepackage{cancel}

\newcommand{\figref}[1]{Fig.~\ref{#1}}

\newcommand{\bea}{\begin{eqnarray}}
\newcommand{\eea}{\end{eqnarray}}
\newcommand{\beaS}{\begin{eqnarray*}}
\newcommand{\eeaS}{\end{eqnarray*}}

\newcommand{\nn}{\nonumber}
\newcommand{\Lag}{{\mathcal L}}

\begin{document}
\title{Nucleon-nucleon scattering with perturbative pions: The uncoupled $\boldsymbol P$-wave channels}

\author{J.~B.~Habashi\footnote{{\tt jbalalhabashi@email.arizona.edu }}}

\affiliation{Department of Physics, University of Arizona, Tucson, Arizona 85721, USA} 

\begin{abstract}
The uncoupled $P$-wave channels of nucleon-nucleon scattering are studied in an effective field theory (EFT) including a perturbative dibaryon field and perturbative pions. Good agreement between EFT results and the Nijmegen partial wave analysis is observed up to a center-of-mass momentum $k \approx 400$ MeV. Using a method that combines renormalization and fitting together, the long-standing convergence problem of EFTs in these channels with perturbative pions, for momenta above the pion mass is addressed from a new perspective.
\end{abstract}
\maketitle

\section{Introduction}
\label{sec.1}

Half a decade after Weinberg's seminal work \cite{Weinberg:1990rz,Weinberg:1991um} on effective field theories (EFTs) of nuclear physics with non-perturbative pions \cite{Ordonez:1992xp,Ordonez:1993tn}, Kaplan, Savage and Wise (KSW) introduced an EFT with a new power counting (PC) that treats pion interactions perturbatively \cite{Kaplan:1996xu,Kaplan:1998tg,Kaplan:1998we}. For a recent review of different PC and progress in nuclear EFTs, see Refs.~\cite{vanKolck:2020plz,Hammer:2019poc}. Similar to Weinberg's PC, i.e., naive dimensional analysis (NDA) \cite{Manohar:1983md,Georgi:1992dw}, the KSW PC requires resummation of the contact $C_{0}$ interaction at leading order (LO) for the $S$-wave channels. This reproduces the bound and virtual state poles seen in nature. Pions, however, were included perturbatively starting at next to leading order (NLO) as justified by empirical evidence from the spin-singlet $S$-wave channel which suggests that the chiral expansion parameter for nucleon-nucleon ($NN$) scattering is roughly $1/3$. This agrees with the theoretical expansion in $Q/\Lambda_{NN}$ where $Q \approx 100$ MeV is a typical low energy scale and $\Lambda_{NN} = 16 \pi\,f_{\pi}^2 / \left( m_{N}\,g_{A}^2 \right) \approx 300$ MeV, with $g_{A} = 1.25$ and the pion decay constant $f_{\pi} \simeq 92$ MeV.

In the $S$-wave channels, pions at NLO only account for half of the effective range $r_{0}$, so $C_{2}$ is promoted to NLO in the KSW PC,\footnote{It has also been shown that with specific renormalization group considerations $C_{2}$ should appear at NLO in the KSW and pionless PC \cite{Kaplan:1996xu,Kaplan:1998tg,Kaplan:1998we,Birse:1998dk,vanKolck:1998bw}.} even though it appears at a higher order in NDA. By using the same argument as in Ref.~\cite{Manohar:1983md}, one can show that changing the PC for $C_{2}$ affects higher order low-energy constants (LECs) in the $S$-wave channels as well. In the other channels,\footnote{Except for the contact interactions of the mixing channel ${}^{3}\!S_{1} - {}^{3}\!D_{1}$, like $C_{2}^{\tiny (SD)}$, that appear at the same order as NDA.} the LECs of four-nucleon contact interactions in the KSW PC have an extra power of $M_{\text{lo}} / M_{\text{hi}}$ relative to NDA. For example, in the $P$-wave channels the LO phase shift is zero, and at NLO and next to next to leading order (NNLO) there are only contributions from the one-pion exchange (OPE) potential; $C_{2}$ in these channels appears at next to next to next to leading order (N$^{3}$LO) \cite{Kaplan:1998we,Fleming:1999ee}.

In the $^{1}\!S_{0}$ channel, the KSW PC converges for momenta up to $\Lambda_{NN} \approx 300$ MeV; however, Fleming {\it et al.} \cite{Fleming:1999ee,Fleming:1999bs} show that in the lower spin-triplet channels the PC does not converge at NNLO. Additionally, Kaplan and Steele \cite{Kaplan:1999qa} argue that the slow convergence problem of the KSW PC above the pion mass is due to higher energy physics and is not related to pion interactions which mostly affect lower energies. 

One way of incorporating correlated high-energy physics is by introducing a dibaryon field. They were first introduced in the context of nuclear EFTs by Kaplan \cite{Kaplan:1996nv} and have since been studied in the $S$-wave channels of NN scattering \cite{Bedaque:1999vb,Soto:2007pg,Soto:2009xy,Long:2013cya,SanchezSanchez:2017tws,Habashi1S0}. These studies were carried out in various energy regimes, both with perturbative and nonperturbative pions.\footnote{For applying a dibaryon as a resonance field in all partial waves of a pionless EFT, see Ref.~\cite{Bedaque:2003wa}.} Comparing the pionless and pionful results of Ref.~\cite{SanchezSanchez:2017tws} for momenta up to $k \approx 400$ MeV strongly suggests that pions can be treated perturbatively in the presence of dibaryon fields. This begs the question: Can pion interactions in other channels be treated perturbatively in the presence of an auxiliary field? Although it might be new in nuclear EFTs, a similar idea was pursued in the context of quantum interactions. In earlier papers \cite{Weinberg:1962hj,Weinberg:1963zza}, Weinberg showed that quantum interactions can be treated perturbatively by introducing a heavy quasi-particle state.

In addition to the work on $S$-wave channels of $NN$ scattering that suggest that pions can be treated perturbatively with a dibaryon field, there is phenomenological motivation, especially for the uncoupled $P$-wave channels, for including a dibaryon field. That is the phase shift in these channels decreases almost linearly at larger momenta, i.e., $\delta \propto - k$; however, promoting the contact $C_{2}$ interaction to NLO gives $\delta \propto k^3$ \cite{Peng:2020nyz}. But with a dibaryon field one can get a tuneable linear function (see Sec.~\ref{sec.4} for further discussion on this subject). 

By adding a dibaryon field, an infinite subset of correlated contact interactions are included which incorporate higher energy physics. One result of this is to push the breakdown scale of the theory to $M_{\text{hi}} \approx 1$ GeV, and the low-energy scale of the theory to be about $M_{\text{lo}} \approx 300$ MeV. Pion interactions also become perturbative by including a dibaryon field. As a result, the PC will differ from NDA.

In the new PC chiral-symmetry-breaking (CSB) contact interactions appear at higher orders after chiral ones. By looking at diagrams and keeping track of divergent terms proportional to $m_{\pi}^{2m}$ we can get an estimate of the PC \cite{Kaplan:1998tg,Kaplan:1998we}. After factoring out $4\pi / m_{N}$, the PC of contact interaction LECs in the Lagrangian Eq.~\eqref{Lagrangian} for $n \geq 0$ and $m \geq 1$ is 
\bea
C_{2n} & \sim & {\cal O}\left(\frac{1}{M_{\text{lo}} M_{\text{hi}}^{2n}}\right) \, , \\
D_{2n+2m}\,m_{\pi}^{2m} & \sim &  {\cal O}\left(\frac{M_{\text{lo}}^{2m-2}}{M_{\text{hi}}^{2n+2m-1}}\right) \sim \left(\frac{M_{\text{lo}}}{M_{\text{hi}}}\right)^{2m-1}\!{\cal O}\left(\frac{1}{M_{\text{lo}}\,M_{\text{hi}}^{2n}}\right) \,. \label{eq.1} 
\eea
For a specific $n$, $C_{2n}$ appears at N$^{2n}$LO. For example, two $C_{0}$ appear at LO of $S$-wave channels and six $C_{2}$ appear at NNLO of $S$- and $P$-wave channels and so on. The CSB LECs with the same powers of momentum as $C_{2n}$s start to appear at one order higher. $m_{\pi}^2$ appears whenever there is a breaking of chiral symmetry; hence, counting divergent terms proportional to $m_{\pi}^{2m}$ suggests that symmetry breaking is down by appropriate powers of $M_{\text{lo}}/M_{\text{hi}}$ each time. For example, the PC of $D_{2n+2m}\,m_{\pi}^{2m}$ suggests that adding $m_{\pi}^{2m}$ is equivalent to adding $2m - 1$ powers of $M_{\text{lo}}/M_{\text{hi}}$ to the normal PC of the same order LECs, i.e., $C_{2n}$s. This observation is useful for determining the PC of CSB LECs of dibaryon fields after finding the PC for their chiral counterparts.

Similar to the PC of NDA and KSW, a cancellation between $C_{0}$ and the analytic part of loop integrals results in a resummation of diagrams. The resummation only happens in the $S$-wave channels, and all other LECs are perturbative, e.g., a $C_{2}$ appears at NNLO in all channels. The first CSB contact interaction $D_{2}$ appears at NLO in the $S$-wave channels, and in all channels $D_{4}$ appears at N$^{3}$LO.

As demonstrated in this paper, the dibaryon field is perturbative and appears at NLO. For the uncoupled $P$-wave channels the PC of the LECs in the Lagrangian Eq.~\eqref{Lagrangian} associated with the dibaryon field is
\bea
g_{1} \sim {\cal O}\left(\frac{1}{M_{\text{hi}}}\right) \ & , &\ h_{1+2m}\,m_{\pi}^{2m} \sim {\cal O}\left(\frac{M_{\text{lo}}^{2m-1}}{M_{\text{hi}}^{2m}}\right) \,, \label{eq.2} \\
\Delta  \sim  {\cal O}\left(\frac{M_{\text{lo}}^2}{M_{\text{hi}}}\right) \ & , &\ \omega_{2m}\,m_{\pi}^{2m}  \sim  {\cal O}\left(\frac{M_{\text{lo}}^{2m+1}}{M_{\text{hi}}^{2m}}\right) \,, \label{eq.3}
\eea
where $g_{1}$ and $\Delta$ are chirally symmetric dibaryon LECs with naive or simple PC, and $h_{1+2m}\,m_{\pi}^{2m}$ and $\omega_{2m}\,m_{\pi}^{2m}$ are CSB dibaryon LECs with the same type of interaction as $g_{1}$ and $\Delta$, respectively. In principle, we should be able to derive this PC for each partial wave by using a nonrelativistic version of the method in Ref.~\cite{Manohar:1983md}, but such a calculation is beyond the scope of this paper. I find that the PC in Eqs.~\eqref{eq.2} and \eqref{eq.3} works at least up to NNLO for the uncoupled $P$-wave channels. The observation about the relation between $m_{\pi}^{2m}$ and $M_{\text{lo}}^{2m-1}/ M_{\text{hi}}^{2m-1}$ for $D_{2n+2m}\,m_{\pi}^{2m}$ is key in deducing the CSB PC in Eqs.~\eqref{eq.2} and \eqref{eq.3}. As we see, the first CSB LECs for $P$ waves appear at NNLO.
 
The rest of the paper is organized as follows. In Sec.~\hyperref[sec.2]{II}, the Lagrangian and PC are discussed in more detail. The $T$ matrix and a new method of renormalization and parameter fitting are explained in Sec.~\ref{sec.3}. In the same section, the NLO and NNLO regulator-independent $T$ matrices are calculated. In Sec.~\ref{sec.4}, the EFT phase shifts are fitted to the Nijmegen partial wave analysis (PWA) \cite{nn-online.org,Stoks:1993tb}. I conclude in Sec.~\ref{sec.5}. Appendix~\ref{AppxA} gives details of the off-shell NLO $T$ matrices, analytic expressions for pion-dibaryon loop integrals and the regulator dependence of LECs. In Appendix~\ref{AppxB}, a second method of renormalization and parameter fitting is presented and discussed.

\section{Lagrangian and power counting}
\label{sec.2}
I consider $NN$ scattering system for which the Lagrangian is invariant under Lorentz transformations (in the form of reparameterization invariance \cite{Luke:1992cs,Jenkins:1990jv}), parity and time-reversal transformations, and conserves baryon number. Similar to NDA \cite{Manohar:1983md,Georgi:1992dw}, the low- and high-energy scales are taken to be $M_{\text{lo}} \approx 300$ MeV and $M_{\text{hi}} \approx 1$ GeV, and pions are needed as explicit degrees of freedom because their mass $m_\pi\simeq 140$ MeV is of order $M_{\text{lo}}$. Spontaneous breaking of chiral SU(2)$_L\times$SU(2)$_R$ symmetry of QCD into the SU(2)$_{V}$ subgroup of isospin \cite{Weinberg:1968de,Weinberg:1990rz,Weinberg:1991um} is a guide for the form of interactions in the EFT, with an isospin-triplet of pions $\pi_a$ ($a=1,2,3$). Chiral-symmetric interactions are included via the chiral covariant derivatives of pion and nucleon fields; however, CSB interactions proportional to quark masses or $m_{\pi}^2$ are constructed using components of $SO(4)$ vectors. Finally, in order to include physics of energies above the pion mass, I introduce an auxiliary odd-parity dibaryon field $\phi_{i,a}^{(s)}$ \cite{Kaplan:1996nv} for each channel with a sign parameter $\eta^{(s)}$. The most general Lagrangian is\footnote{I assume that only nucleon fields interact with pions and not the dibaryon field. For a dibaryon field interacting with pions, see Refs.~\cite{Soto:2007pg,Soto:2009xy,Long:2013cya}.}
\bea
\Lag & = &  N^{\dagger} \left[ i \partial_{0} + \frac{\vec{\nabla}^2}{2m_N} 
- \frac{g_A}{2f_\pi} \,\tau_{a} 
\left(\vec{\sigma}\cdot\vec{\nabla}\pi_{a}\right)\right] N + \frac{1}{2}\, \pi_a\left(\partial_0^2-\vec{\nabla}^2+m_{\pi}^{2}\right)\pi_a \nn \\ 
&& +\ \eta^{(s)}\phi_{i, a}^{(s) \dagger} \left[ i\partial_{0} + \frac{\vec{\nabla}^2}{4 m_N} - \left(\Delta^{(s)}+\omega_{2}^{(s)}\,m_{\pi}^2\right) \right] \phi_{i, a}^{(s)} + \sqrt{\frac{4\pi}{m_N}}\,\left(g_{1}^{(s)}+h_{3}^{(s)}\,m_{\pi}^2\right) \left(\phi_{i, a}^{(s) \dagger} N^{T} P_{i,a}^{(s)}N + \mathrm{H.c.}\right) \nn \\ 
&& -\ \frac{4 \pi}{m_N} C_{2}^{(s)} \left(N^{T}P_{i,a}^{(s)}
N\right)^{\dagger} \left(N^{T}P_{i,a}^{(s)} N \right) +\ldots\,,
\label{Lagrangian}
\eea
where $s = {}^{1}\!P_{1}, {}^{3}\!P_{0}, {}^{3}\!P_{1}$, and the sum over the vector index $i$ and isospin index $a$ are implicit. LECs are also labeled for each channel separately. In the above Lagrangian, $\vec{\sigma}$ ($\tau_a$) are the Pauli matrices in spin (isospin)
space, 
\bea
P_{i}^{\tiny ({}^{1}\!P_{1})} & = & \frac{\sqrt{3}}{2 \sqrt{2}} \left(\frac{-i\overleftrightarrow{\nabla}_{i}}{2}\right)\left( i \sigma_{2}\right)\left( i \tau_{2}\right) \, , \\
P_{a}^{\tiny ({}^{3}\!P_{0})} & = &  \frac{1}{2\sqrt{2}}\left(\frac{- i\,\overleftrightarrow{\nabla}_{i}}{2}\right)\left( i \sigma_{2}\,\sigma_{i}\right)\left( i \tau_{2}\,\tau_{a}\right) \, , \\
P_{i,a}^{\tiny ({}^{3}\!P_{1})} & = & \frac{\sqrt{3}}{4} \, \epsilon_{ijk} \left(\frac{-i\overleftrightarrow{\nabla}_{j}}{2}\right)\left( i \sigma_{2}\,\sigma_{k}\right)\left( i \tau_{2}\,\tau_{a}\right)
\label{projector}
\eea
are the projection operators on respectively the ${}^{1}\!P_{1}$, ${}^{3}\!P_{0}$ and ${}^{3}\!P_{1}$ channels \cite{Fleming:1999ee,Fleming:1999bs}, with $\overleftrightarrow{\nabla}\equiv\overrightarrow{\nabla}-\overleftarrow{\nabla}$. The ``$\ldots$" indicates higher order terms with more fields, derivatives and powers of $m_{\pi}^2$. For a nucleon-dibaryon interaction with $\ell$ derivatives that appears at NLO we need at least $\ell + 1$ dibaryon LECs to renormalize the loop integrals at higher orders. For the $P$ waves with $\ell = 1$, we need two LECs, which are present as $\Delta$ and $g$, and there is no need for higher derivative dibaryon interactions, as their effects are accounted for by higher order contact terms.

In contrast to the usual notation in most of the literature, $4\pi/m_N$ has been factored out of the LECs, and the convention for definitions of the $T$ matrix and potential throughout the paper is
\bea
\overline{T} \equiv \frac{m_{N}\,T}{4 \pi} \qquad , \qquad \overline{V} \equiv \frac{m_{N}\,V}{4 \pi} \, .
\label{TVconv}
\eea
A typical LEC $\mathtt{g}$ and consequently observables like the $T$ matrix and phase shift are expanded at each order of the perturbation
\bea
\mathtt{g} & = & \mathtt{g}^{(0)} + \mathtt{g}^{(1)} + \mathtt{g}^{(2)} + \ldots \, ,\\
\overline{T} & = & \overline{T}^{(0)} + \overline{T}^{(1)} + \overline{T}^{(2)} + \ldots \, , \label{expT} \\
\delta & = & \delta^{(0)} + \delta^{(1)} + \delta^{(2)} + \ldots \label{expPhase} \, ,
\eea
where the superscripts $n = 0, 1, 2, \ldots$ indicate LO, NLO, NNLO, and so on. Each channel has its own LECs, but the channel superscripts $s$ will be dropped throughout this paper for simplicity. According to the PC, the LECs and interactions appearing at each order are
\bea
\textrm{LO} & : & ---,\nn\\
\textrm{NLO} & : & \Delta^{(0)},\ g_{1}^{(1)} ,\ \vec{\pi}, \\
\textrm{NNLO} & : & \Delta^{(1)},\ \omega_{2}^{(1)} m_{\pi}^2,\ g_{1}^{(2)},\ h_{3}^{(2)} m_{\pi}^2,\ C_{2}^{(2)},\ \vec{\pi}\,\vec{\pi}, \nn
\eea
where $\vec{\pi}\vec{\pi}$ means iteration of the OPE potential. Radiation and soft pion interactions are not considered in the up to NNLO calculations of this paper. Thus, similar to Ref.~\cite{Fleming:1999ee} for the $P$ waves, I include only iterations of the OPE potential. These iterations contain infrared enhancements \cite{Weinberg:1990rz,Weinberg:1991um} relative to multiple-pion-exchange potentials with the same number of pion fields \cite{Kaiser:1997mw}. Note that the residual mass of the dibaryon always comes with a factor of the nucleon mass $m_{N}$, and therefore $\Delta^{(0)}$ is a LO LEC with the same size as the kinetic term, but the smallness of the nucleon-dibaryon coupling $g_{1}$ means that it first appears at NLO. 

There is no difference in the Feynman rules of chiral and CSB LECs with the same number of derivatives or momenta that appear at the same order in the perturbation. Thus, I define
\bea
\tilde{\Delta}^{(1)} & \equiv & \Delta^{(1)} + \omega_{2}^{(1)}\,m_{\pi}^2 \, \\
\tilde{g}_{1}^{(2)} & \equiv & g_{1}^{(2)} + h_{3}^{(2)}\,m_{\pi}^2 \, ,
\eea
which are used in the NNLO calculations. Whenever a new LEC appears at a given order of perturbation theory, e.g., $\omega_{2}^{(1)}$ and $h_{3}^{(2)}$, an additional fitting parameter is needed as well. The NLO LECs and NNLO $C_{2}$ do not run with the regulator, and I give an estimate of their size in terms of $M_{\text{lo}}$ and $M_{\text{hi}}$ in Sec.~\ref{sec.4}.

In this paper, I renormalize in two different ways. In the next section, I maximally exploit the nonlocality of the dibaryon field by absorbing a part of the nonanalytic effects of pion interactions in its NNLO LECs. In the second method detailed in Appendix~\ref{AppxB}, nonanalytic pion effects get absorbed in a redefinition of dibaryon NLO LECs by refitting them at NNLO. Although this may work in special cases and at lower orders of the perturbation, we see that nonanalytic effects of pion interactions, both chiral and CSB parts, can have large effects on LECs at higher orders of the perturbation, as has been observed in Refs.~\cite{Fleming:1999ee,Fleming:1999bs}.

\section{$T$ matrix and Renormalization}
\label{sec.3}
The $T$ matrix of nonrelativistic quantum systems has been studied extensively; for example, see Ref.~\cite{TaylorScattering}. I am specifically interested in the $T$ matrix that is directly related to the scattering amplitude via Feynman diagrams in an EFT.\footnote{Note that for the scattering amplitude $\mathcal{A}$ we have $\mathcal{A} = - \overline{T}$.} Therefore, for a nonrelativistic EFT, I use the Lippmann-Schwinger equation (LSE) for a systematic expansion of $T$ in a perturbative scheme. The off-shell LSE for the $T$ matrix in the momentum space is
\bea
T(\vec{p}\,', \vec{p}, k) & = & V(\vec{p}\,', \vec{p}) + \int \frac{d^3 q}{(2 \pi)^{3}} V(\vec{p}\,', \vec{q}) \;G(\vec{q},k) \;T(\vec{q}, \vec{p})\ , \qquad
\eea
where $k = \sqrt{m_{N} E}$ and
\bea
G \left( \vec{q}, k \right) = \frac{m_{N}}{k^2 - q^2 + i\, \epsilon} \, .
\eea
is the nonrelativistic Schr\"odinger propagator. The potential $V(\vec{p}\,', \vec{p})$ is given by tree level diagrams in the EFT. The total spin $s$ and the total angular momentum $j$ are conserved, and if I label the angular momentum between incoming and outgoing particles by $\ell$ and $\ell'$ respectively, the projected off-shell $T$ matrix with the convention in Eq.~\eqref{TVconv} is
\bea
\overline{T}_{\ell, \ell'}\left(p', p\right) & = & \overline{V}_{\ell, \ell'}(p', p) + \frac{2}{\pi} \sum_{\ell''}\int^{\Lambda} d q\,q^{2} \frac{\overline{V}_{\ell, \ell''}(p', q)\,\overline{T}_{\ell'', \ell'}(q, p)}{k^2 - q^2 + i \epsilon} \, , \qquad
\label{ProjectedLSE}
\eea
where the angular momenta $\ell$, $\ell'$ and $\ell''$ run over $j-s$ to $j+s$. For the uncoupled channels, $\ell = \ell' = \ell''$ and I ignore the summation. In the above equation, the sharp-cutoff regulator is only for the magnitude of 3-momentum, and there is no breaking of rotational invariance. 

For the sharp cutoff with $\Lambda\to\infty$ limit, Phillips {\it et al.} \cite{Phillips:1998uy} show that results of loop integration are equivalent to the power divergence subtraction PDS) scheme of Refs.~\cite{Kaplan:1998tg,Kaplan:1998we} if poles in dimensions other than $\mathrm{D} = 4$ are subtracted. In the sharp-cutoff scheme of this paper, I keep only the divergent and finite terms in loop diagrams. Although terms that vanish as $\Lambda\to\infty$ give insight into the size of the theoretical error and the next order LECs, they are omitted because they are not important for the purpose of renormalization. 

In order to find the $T$ matrix at each order in perturbation theory, I calculate the tree level off-shell $T$ matrix or potential, and then replace the $T$ matrix in the integrand of Eq.~\eqref{ProjectedLSE} with the potential to get the $T$ matrix of next order and so on. The phase shift and $T$ matrix are related to each other through the unitarity condition of the $S$ matrix
\bea
S & = & e^{2 i\,\delta} = 1 - 2\,i\,k\,\overline{T} \, .
\eea
For the $P$-wave channels, the LO phase shift $\delta^{(0)} = 0$ because the $T$ matrix vanishes at this order. Therefore after using expansions in Eqs.~\eqref{expT} and \eqref{expPhase}, relations between phase shifts and $T$ matrices of higher orders are
\bea
\delta^{(1)} & = & -k\,\overline{T}^{(1)} \label{NLOphase}\\
\delta^{(2)} & = & -k\,\overline{T}^{(2)} - i \,k^{2}\,{\overline{T}^{(1)}}^2 \label{NNLOphase}\\
&\vdots & \nn
\eea
Note that the phase shift is real and there is no imaginary piece in Eq.~\eqref{NLOphase}. Also, the imaginary part of NNLO $T$ matrix cancels the last term in Eq.~\eqref{NNLOphase}. This is another way of checking the unitarity condition of the $S$ matrix at each order in the perturbation.

\subsection{The NLO and NNLO $T$ matrices}
\begin{figure}
\centering
\includegraphics[trim={7.88cm 24.25cm 7.89cm 2.15cm},clip]{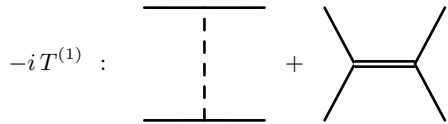}
\caption{Feynman diagrams corresponding to the NLO $T$ matrix. The OPE potential contribution is shown on the left and the dibaryon contribution on the right. The nucleon-dibaryon coupling is $g_{1}^{(1)}$.}
\label{fig.1}
\end{figure}
In the three channels I consider in this paper, the $T$ matrix and the renormalization procedure are independent of specific channel projections. Hence, at NLO there are OPE and dibaryon interactions shown in \figref{fig.1}. The on-shell $T$ matrix corresponding to these diagrams is
\bea
\overline{T}^{(1)}\left(k, m_{\pi}^2\right) = \overline{T}^{(1)}_{\pi}\left(k, m_{\pi}^2\right) + \frac{\eta  \,m_{N}\,{g_{1}^{(1)}}^2\,k^2}{k^{2}  - m_{N}\,\Delta^{(0)}}\, .
\eea
where $\overline{T}_{\pi}^{(1)}\left(k, m_{\pi}^2\right)$, which is the on-shell tree level OPE $T$ matrix for each channel, is given in the next section by taking the on-shell limit of the off-shell NLO $T$ matrix in Appendix~\ref{AppxA}.

\begin{figure}
\centering
\includegraphics[trim={3.98cm 22.45cm 4.15cm 2.15cm},clip]{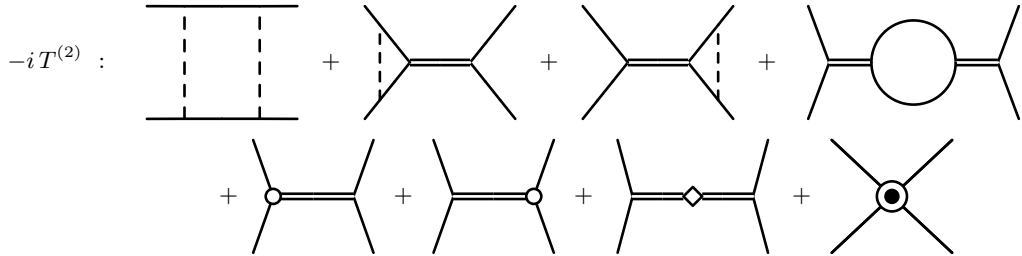}
\caption{Feynman diagrams corresponding to the NNLO $T$ matrix. The first line involves NLO vertices and the second line involves NNLO vertices. The circle, diamond and double circle represent $\tilde{g}_{1}^{(2)}$, $\tilde{\Delta}^{(1)}$ and $C_{2}^{(2)}$ respectively.}
\label{fig.2}
\end{figure}

Feynman diagrams corresponding to the total NNLO $T$ matrix are shown in \figref{fig.2}. The two-pion ladder or box diagram is not divergent for the $P$ waves; however, the cross pion-dibaryon diagrams are linearly divergent (except for one channel), and the two-dibaryon diagram with a loop is divergent as well. While it is possible to derive analytic expressions of the divergent and pionless parts of the NNLO $T$ matrix, I calculate the finite and real part of the first diagram in \figref{fig.2} numerically by the method described in the previous section.\footnote{An alternative numerical calculation of the box diagram from the OPE potential can be found in Ref.~\cite{Kaiser:1997mw}.} The sum of the finite and real parts of the first three diagrams in \figref{fig.2} is 
\bea
\mathcal{R}^{(2)}\!\left(k, g_{1}^{(1)}, \Delta^{(0)}, m_{\pi}^2\right)\!\!&=&\!\! \mathtt{Re}\left[\overline{T}_{\pi\pi}^{(2)}\left(k, m_{\pi}\right)\right] + \frac{2\,\eta  \,m_{N}\,{g_{1}^{(1)}}^2\,k^2}{k^{2}  - m_{N}\,\Delta^{(0)}}\, \mathtt{Re}\left[ I_{\pi d}^{[\mathtt{fin}]} \right]\, , \qquad
\label{realNNLOT}
\eea
where $\overline{T}_{\pi\pi}^{(2)}\left(k, m_{\pi}\right)$ is the on-shell $T$ matrix of the box diagram from iteration of the OPE potential, and $I_{\pi d}^{[\mathtt{fin}]}$ is the dimensionless finite part of the loop integral in cross pion-dibaryon diagrams given in Appendix~\ref{AppxA}. Then, the $T$ matrix at NNLO can be expressed as
\bea
\overline{T}^{(2)}\!\!\left(k, m_{\pi}^2\right) \!\!&=& - i\,k\;\left[\overline{T}^{(1)}\left(k, m_{\pi}^2\right)\right]^2 + \mathcal{R}^{(2)}\left(k, g_{1}^{(1)}, \Delta^{(0)}, m_{\pi}^2\right) + \frac{2\,\eta\,m_{N}\,{g_{1}^{(1)}}^2\,k^2}{k^{2} - m_{N}\,\Delta^{(0)}}\,\biggl[ I_{\pi d}^{[\mathtt{div}]} - \eta\, m_{N}\,{g_{1}^{(1)}}^2 \frac{L_{1}}{2} + \frac{\tilde{g}_{1}^{(2)}}{g_{1}^{(1)}}\biggr] \nn\\
&& +\ C_{2}^{(2)}\,k^2 - \frac{m_{N}^2\,{g_{1}^{(1)}}^2\,k^2}{\left(k^{2}  - m_{N}\,\Delta^{(0)}\right)^2} \biggl[ {g_{1}^{(1)}}^2\!\!\left(L_{3} + m_{N}\,\Delta^{(0)}\,L_{1} \right) - \eta\,\tilde{\Delta}^{(1)} \biggr]   \, , \label{TNNLO1}
\eea
where $I_{\pi d}^{[\mathtt{div}]}$ is the dimensionless divergent part of the loop integral in cross pion-dibaryon diagrams given in Appendix~\ref{AppxA}, and for a nonzero integer $m$ and with a sharp-cutoff regulator, $L_{m}$ is
\bea
L_{m} = \frac{2}{\pi} \frac{\Lambda^m}{m} \, .
\label{Lm}
\eea

A few remarks: First, from Eqs.~\eqref{Idpi1P1}--\eqref{Idpi3P1} in Appendix~\ref{AppxA}, $I_{\pi d}^{[\mathtt{div}]}$ is not proportional to $m_{\pi}^2$, so there is no need for $D_{4}\,m_{\pi}^2$ at NNLO; however, the central values of the ``chiral" NLO LECs will change due to the inclusion of $\omega_{2}^{(1)}$ and $h_{3}^{(2)}$ in $\tilde{\Delta}^{(1)}$ and $\tilde{g}_{1}^{(2)}$ respectively. But I can renormalize in a way that the central values of fitted chiral NLO plus CSB NNLO parameters stay unchanged. Second, all divergent terms in Eq.~\eqref{TNNLO1} can be absorbed by the NNLO expansions of NLO LECs, and therefore $C_{2}^{(2)}$ is regulator independent. I find its value by fitting the result of the up to NNLO EFT phase shift to data. Finally, all imaginary parts of the NNLO $T$ matrix come from diagrams in the first line of \figref{fig.2}, and one can show numerically that their sum is the same as the imaginary term in Eq.~\eqref{TNNLO1}, as is required by the unitarity condition of the $S$ matrix.

For renormalization, I use a new method that keeps central fitted values of chiral NLO plus CSB NNLO parameters unchanged. There are two NLO LECs and two data points are needed at momenta $k = \{k_{1}, k_{2}\}$ to fix them. These fitted values are directly related to the experimental values or a model that describes the phase shift (like PWA) at those momenta. Therefore, in the up to NLO EFT I can attribute these fitted values to chiral parts, but at NNLO adding $\omega_{2}^{(1)}$ and $h_{3}^{(2)}$ means a portion of these fitted values are coming from CSB terms. 

Since I want to keep central values of the linear combination of chiral NLO and CSB NNLO terms unchanged at momenta used in the NLO fit, the real part of the NNLO $T$ matrix in Eq.~\eqref{TNNLO1} should contribute to the total up-to-NNLO $T$ matrix ``only" as a renormalized CSB value. Hence, I have
\bea
\mathtt{Re}\left[\overline{T}^{(2)}\left(k_{1,2}, m_{\pi}^2\right)\right] & = & \frac{\eta\,m_{N}\,\gamma^{(2)}\,m_{\pi}^2\,k_{1,2}^2}{k_{1,2}^{2} - m_{N}\,\Delta^{(0)}} + \frac{\eta\,m_{N}^2\,{g_{1}^{(1)}}^2\,\theta^{(1)}\,m_{\pi}^2\,k_{1,2}^2}{\left(k_{1,2}^{2}  - m_{N}\,\Delta^{(0)}\right)^2} \, ,\qquad \label{2conditions}
\eea
where $\theta^{(1)}$ and $\gamma^{(2)}$ are renormalized values related to $\omega_{2}^{(1)}$and $h_{3}^{(2)}$ respectively. The above equation is an abbreviation for two conditions\footnote{In a pionless theory, the right hand side of Eq.~\eqref{2conditions} is zero which is similar to conditions put on pionless theories by keeping the experimental values of the effective range parameters fixed.} at $k_{1}$ and $k_{2}$. From these two conditions, I can find $\tilde{g}_{1}^{(2)}$ and $\tilde{\Delta}^{(1)}$, given in Appendix~\ref{AppxA}, in terms of NLO fitted values, the cutoff $\Lambda$, $C_{2}^{(2)}$ and complicated nonanalytic functions of $m_{\pi}$ \cite{Kaplan:1999qa,Rupak:1999aa}. With the conditions in Eq.~\eqref{2conditions}, the renormalized NNLO $T$ matrix is
\bea
\overline{T}^{(2)}\left(k, m_{\pi}^2\right) &=& - i\,k\;\left[\overline{T}^{(1)}\left(k, m_{\pi}^2\right)\right]^2 +  
\frac{\eta\,m_{N}\,\gamma^{(2)}\,m_{\pi}^2\,k^2}{k^{2} - m_{N}\,\Delta^{(0)}} + \frac{\eta\,m_{N}^2\,{g_{1}^{(1)}}^2\,\theta^{(1)}\,m_{\pi}^2\,k^2}{\left(k^{2}  - m_{N}\,\Delta^{(0)}\right)^2} + \mathcal{R}^{(2)}\left(k, g_{1}^{(1)}, \Delta^{(0)}, m_{\pi}^2\right)
\nn \\
&& -\ \frac{\left(k^2 - k_{2}^2\right)}{\left(k_{1}^2 - k_{2}^2\right)} \frac{\left(k_{1}^{2} - m_{N}\,\Delta^{(0)}\right)^2}{\left(k^{2} - m_{N}\,\Delta^{(0)}\right)^2}\,\frac{k^2}{k_{1}^2}\,\mathcal{R}^{(2)}_{k_{1}} + \frac{\left(k^2 - k_{1}^2\right)}{\left(k_{1}^2 - k_{2}^2\right)} \frac{\left(k_{2}^{2} - m_{N}\,\Delta^{(0)}\right)^2}{\left(k^{2} - m_{N}\,\Delta^{(0)}\right)^2}\,\frac{k^2}{k_{2}^2} \,\mathcal{R}^{(2)}_{k_{2}} \nn\\
&& +\ \frac{\left(k^2 - k_{1}^2\right)\left(k^2 - k_{2}^2\right)}{\left(k^{2} - m_{N}\,\Delta^{(0)}\right)^2}\,C_{2}^{(2)}\,k^2
\, , \label{finalNNLOT}
\eea
with $\mathcal{R}^{(2)}_{k_{1,2}} \equiv \mathcal{R}^{(2)}\left(k_{1,2}, g_{1}^{(1)}, \Delta^{(0)}, m_{\pi}^2\right)$. 
The total phase shift up to NNLO from Eqs.~\eqref{NLOphase} and \eqref{NNLOphase} is
\bea
-\frac{\delta^{(t)}}{k} &=& -\frac{1}{k}\left(\delta^{(1)} + \delta^{(2)}\right) = \overline{T}^{(1)}_{\pi}\left(k, m_{\pi}^2\right) + \frac{\eta  \,m_{N}\,\bar{g}{_{1}^{(1)}}^2\,k^2}{k^{2}  - m_{N}\,\bar{\Delta}^{(0)}} + \mathcal{R}^{(2)}\left(k, g_{1}^{(1)}, \Delta^{(0)}, m_{\pi}^2\right) + \frac{\left(k^2 - k_{1}^2\right)\left(k^2 - k_{2}^2\right)}{\left(k^{2} - m_{N}\,\Delta^{(0)}\right)^2}\,C_{2}^{(2)}\,k^2 
\nn \\
&& -\ \frac{\left(k^2 - k_{2}^2\right)}{\left(k_{1}^2 - k_{2}^2\right)} \frac{\left(k_{1}^{2} - m_{N}\,\Delta^{(0)}\right)^2}{\left(k^{2} - m_{N}\,\Delta^{(0)}\right)^2}\,\frac{k^2}{k_{1}^2}\,\mathcal{R}^{(2)}_{k_{1}} + \frac{\left(k^2 - k_{1}^2\right)}{\left(k_{1}^2 - k_{2}^2\right)} \frac{\left(k_{2}^{2} - m_{N}\,\Delta^{(0)}\right)^2}{\left(k^{2} - m_{N}\,\Delta^{(0)}\right)^2}\,\frac{k^2}{k_{2}^2} \,\mathcal{R}^{(2)}_{k_{2}} \, ,\qquad \label{totalPhase}
\eea
where the finite $\bar{\Delta}^{(0)}$ and $\bar{g}_{1}^{(1)\,2}$ are
\bea
\bar{\Delta}^{(0)} & \equiv & \Delta^{(0)} + \theta^{(1)}\,m_{\pi}^2 \, , \label{deltabar}\\
\bar{g}{_{1}^{(1)}}^2 & \equiv & {g_{1}^{(1)}}^2 + \gamma^{(2)}\,m_{\pi}^2 \, . \label{g1bar}
\eea
This method of renormalization uses data points instead of an interval of data, so the number of data points I need at each order in the perturbation is equivalent to the number of new LECs at that order. For example, at NLO I have two LECs and two data points are needed at $k = \{k_{1}, k_{2}\}$. At NNLO, in principle I need three additional data points to determine $\theta^{(1)}$, $\gamma^{(2)}$ and $C_{2}^{(2)}$; however, as we can see in Eqs.~\eqref{deltabar} and\eqref{g1bar}, it is impossible to distinguish between chiral and CSB fitting parameters from one scattering process with a fixed value of $m_{\pi}^2$. Therefore, I only need one more data point at $k = k_{3}$ to determine only $C_{2}^{(2)}$, and the central values of $\bar{\Delta}^{(0)}$ and $\bar{g}_{1}^{(1)\,2}$ are kept fixed to the NLO fitted values. 

Note that keeping the NLO fitted values fixed does not mean that $\bar{\Delta}^{(0)} = \Delta^{(0)}$ or $\bar{g}_{1}^{(1)\,2} = g_{1}^{(1)\,2}$. It means that the main portion of the actual values that I fit at NLO are chiral, and a small and perturbative part of them are due to CSB parts, which I could not resolve at NLO. At NNLO, however, effects of these CSB parts become manifest to order $m_{\pi}^2$, and they can be distinguished via $\theta^{(1)}$ and $\gamma^{(2)}$, although their combinations with chiral parts still have the same fitted values as at NLO. In order to find values for $\theta^{(1)}$ and $\gamma^{(2)}$ new data points are needed either from lattice calculations of the same process with a different value of $m_{\pi}^2$ or another $m_{\pi}$-sensitive process. Since $\Delta^{(0)}$ and $g_{1}^{(1)\,2}$ in Eq.~\eqref{totalPhase} appear in NNLO terms, I can replace them with $\bar{\Delta}^{(0)}$ and $\bar{g}_{1}^{(1)\,2}$ and use the same fitted NLO values when I fit the total phase shift to data. A different method of renormalization is explained in Appendix~\ref{AppxB}.

Forms of the last three terms in Eq.~\eqref{totalPhase} are specific combinations of terms in Eq.~\eqref{TNNLO1}. Although these forms ensure that two conditions in Eq.~\eqref{2conditions} are fulfilled, their overall effect is more than just that. They counteract effects of nonanalytic terms in pion interactions that arise in $\mathcal{R}^{(2)}\left(k, g_{1}^{(1)}, \Delta^{(0)}, m_{\pi}^2\right)$. It is remarkable that simple functions like these rectify effects of a complicated function like $\mathcal{R}^{(2)}$ in almost the entire region of validity of the EFT.


\section{Results}
\label{sec.4}
In order to fit EFT results, three data points are needed for three LECs $\Delta$, $g_{1}$ and $C_{2}$. Higher order contributions to these LECs do not change number of parameters. I choose two sets of data points for each channel in order to show that final results have not been optimized by a specific choice of momenta. Since $\Delta^{(0)}$, $g_{1}^{(1)}$ and $C_{2}^{(2)}$ are finite, their order of magnitude can be estimated according to the PC and the sizes of $M_{\text{lo}} \sim 300$ MeV and $M_{\text{hi}} \sim 1$ GeV:
\bea
\sqrt{m_{N}\,\Delta^{(0)}} &\sim & M_{\text{lo}} \sim 300 \textrm{ MeV}  \nn \\ 
g_{1}^{(1)} &\sim & \frac{1}{M_{\text{hi}}} \sim 10^{-3} \textrm{ MeV}^{-1} \label{estimationLEC} \\ 
C_{2}^{(2)} &\sim & \frac{1}{M_{\text{lo}}\,M_{\text{hi}}^2} \sim 10^{-9}\!-\!10^{-8} \textrm{ MeV}^{-3} \; . \quad \nn
\eea
In order to get values of $\Delta^{(0)}$ and $g_{1}^{(1)}$, I choose two momenta that give the best fit to data. For the third data point needed for $C_{2}^{(2)}$, , I can choose a data point which gives the best result at NNLO.
The total up to NNLO EFT phase shift is given in Eq.~\eqref{totalPhase}, , and the phase shift data is from the Nijmegen PWA \cite{nn-online.org,Stoks:1993tb} for laboratory energies $E_{\text{lab}} \leq 350$ MeV.

As has been shown by Fleming {\it et al.} \cite{Fleming:1999ee,Fleming:1999bs} and Pav\'on Valderrama {\it et al.} \cite{PavonValderrama:2016lqn}, while the phase shift from iterating the OPE potential in the ${}^{1}\!P_{1}$ channel converges quickly, it does a poor job of describing the data above the pion mass. Adding a dibaryon field goes a long way in solving the convergence-to-data problem. The on-shell $T$ matrix in the ${}^{1}\!P_{1}$ channel at NLO is (see Appendix~\ref{AppxA})
\bea
\overline{T}^{(1)\tiny({}^{1}\!P_{1})}\!\left(k, m_{\pi}^2\right) \!\!& = & \frac{\eta\,m_{N}\,{g_{1}^{(1)}}^2\,k^2}{k^{2}  - m_{N}\,\Delta^{(0)}} + \frac{1}{\Lambda_{NN}}\biggl[ -\frac{3\,m_{\pi}^2}{2\,k^2} + \frac{3\,m_{\pi}^2\left(m_{\pi}^2 + 2\,k^2\right)}{8\,k^4}\,\ln\left(1 + \frac{4\,k^2}{m_{\pi}^2}\right) \biggr] \, . \qquad
\eea
\begin{table*}[t]
\centering
\begin{tabular*}{\textwidth}{@{\extracolsep{\fill} } c  c  c  c  c  c  c }
\hline\hline 
${}^{1}\!P_{1}$ & $\{k_{1}, k_{2}, k_{3}\}$(MeV) & \; $g_{1}^{(1)}$ (MeV$^{-1}$)\; & \; $\Delta^{(0)}$ (MeV) \; & \; $C_{2}^{(2)}$ (MeV$^{-3}$) \; & \; $\sqrt{m_{N} |\Delta^{(0)}|}$ (MeV) \; & \; $\eta$ \;\\ \hline
Fit 1 & 350, 400, 300 & 0.00112 & -97.8 & -3.8$\times 10^{-9}$ & 303.0 & +1\\ 
Fit 2 & 310, 370, 280 & 0.00123 & -149.2 & -7.7$\times 10^{-9}$ & 374.3 & +1 
\\\hline\hline
\end{tabular*}
\caption{
Results of fitting up to NNLO LECs in the ${}^{1}\!P_{1}$ channel to the Nijmegen PWA phase shift for two sets of data points at momenta $\{k_{1}, k_{2}, k_{3}\}$. After fitting NLO constants $g_{1}^{(1)}$ and $\Delta^{(0)}$ to the data points $\{k_{1}, k_{2}\}$, $k_{3}$ fixes the $C_{2}^{(2)}$ at NNLO. Note that the fitted value of $\bar{\Delta}^{(0)}$ at NNLO is the same as the fitted value of $\Delta^{(0)}$ at NLO.} 
\label{tbl.1}
\end{table*}
\begin{figure*}[t]
\centering
\includegraphics[scale=0.21]{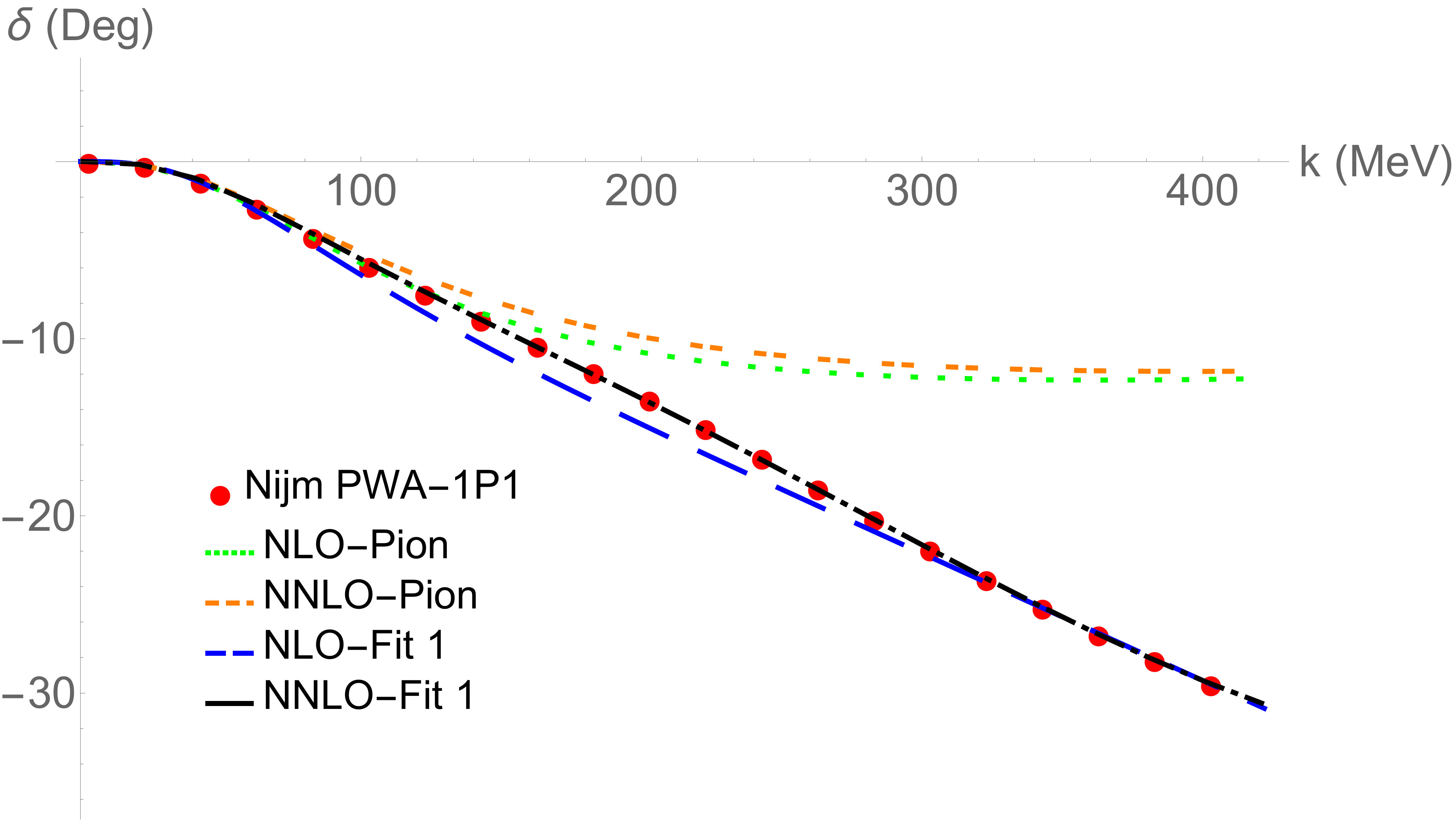}
\includegraphics[scale=0.21]{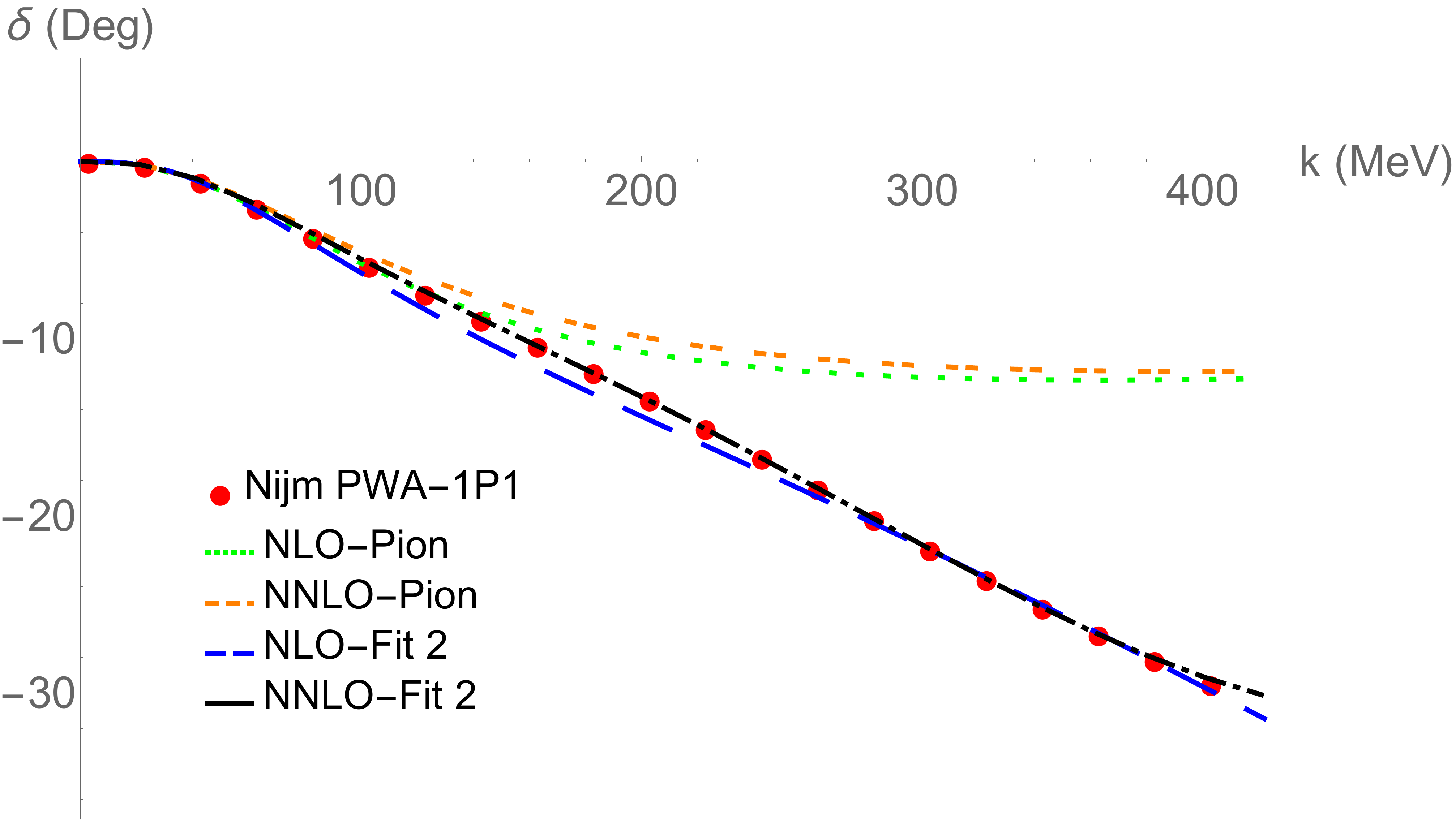}
\caption{The Nijmegen PWA (red circles) vs NLO (blue long-dashed line) and NNLO (black short-dashed line) EFT results for the ${}^{1}\!P_{1}$ channel. Also shown are results from pion-only contributions, which represent the KSW PC, at NLO (green dotted line) and NNLO (orange double-dotted line). I use momentum sets given in Table.~\ref{tbl.1} for fitting and graph results on the left for fit 1, and on the right for fit 2.}
\label{fig.3}
\end{figure*}

Results of fitting the above $T$ matrix to two data points $\{k_{1}, k_{2}\}$ are given in Table.~\ref{tbl.1} and the corresponding phase shift is plotted in \figref{fig.3} (blue long-dashed lines). Even at NLO, an improvement relative to perturbative OPE potential for momenta above $k\approx 200$ MeV is evident. I use Eqs.~\eqref{realNNLOT} and \eqref{finalNNLOT} to fit the NNLO EFT calculated phase shift to the Nijmegen PWA at the third data point, $k = k_{3}$. Results are given in Table.~\ref{tbl.1} and plotted in \figref{fig.3} (black short-dashed lines). Clearly NNLO results are in an even better agreement with data than those at NLO.

Another way of checking the PC is by looking at fitted finite values of LECs. Size of fitted parameters in Table.~\ref{tbl.1} are close to the estimation from the PC in Eq.~\eqref{estimationLEC}. With a negative $\Delta^{(0)}$, the denominator of the NLO $T$ matrix vanishes for an imaginary momentum $\sqrt{m_{N} |\Delta^{(0)}|}$. This imaginary momentum cannot be an ``imaginary" pole corresponding to a bound and/or virtual state, because an imaginary pole appears when there is a resummation in loop integrals and the unitary part of the loop integrals are on equal footing to the imaginary pole, which is not the case for the perturbative approach here. It is possible that this imaginary momentum corresponds to an imaginary zero of amplitude, which is not visible in the real phase shift or in $k\cot\delta$. 

The interesting $k^{2} \gg m_{N}\,\Delta^{(0)},\,m_{\pi}^2$ limit of the LO $T$ matrix is
\bea
\lim\limits_{k^{2} \gg m_{N}\,\Delta^{(0)},\,m_{\pi}^2} \overline{T}^{(1)\tiny(^{1}\!P_{1})} = \eta\,m_{N}\,{g_{1}^{(1)}}^2 + \ldots \, ,
\label{klimit1P1}
\eea
where $\eta = + 1$ for this channel means the phase shift in Eq.~\eqref{NLOphase} decreases linearly at larger momenta. This is a feature coming from the dibaryon field, not the contact interactions. The same feature holds for the  ${}^{3}\!P_{0}$ and  ${}^{3}\!P_{1}$ channels too.

The lower spin-triplet channels are real challenges for the KSW PC at NNLO because of the large disagreement with data \cite{Fleming:1999ee}. These channels have thus been subject of many studies; for example, see Refs.~\cite{Nogga:2005hy,Long:2011qx,Long:2011xw,Wu:2018lai,Kaplan:2019znu,Peng:2020nyz}. The ${}^{3}\!P_{0}$ channel is a special case because unlike the other uncoupled $P$-wave channels its phase shift does not decrease or increase monotonically and it passes zero at $k \approx 300$ MeV similar to the ${}^{1}\!S_{0}$ channel. In contrast, however, the overall size of the phase shift is not large enough to support the idea of a pole in the $T$ matrix. 

For the ${}^{3}\!P_{0}$ channel, the total on-shell $T$ matrix at NLO can be extracted from the off-shell $T$ matrix in the diagrams in \figref{fig.1} (see Appendix~\ref{AppxA}) 
\bea
\overline{T}^{(1)\tiny({}^{3}\!P_{0})}\!\left(k, m_{\pi}^2\right) \!\!&=&\!\! \frac{\eta\,m_{N}\,{g_{1}^{(1)}}^2\,k^2}{k^{2}  - m_{N}\,\Delta^{(0)}} + \frac{1}{\Lambda_{NN}}\biggl[ -1 + \frac{m_{\pi}^2}{4\,k^2}\,\ln\left(1 + \frac{4\,k^2}{m_{\pi}^2}\right) \biggr] \, . \qquad
\eea
The $k^{2}$ term in the numerator of dibaryon part is due to the $P$-wave nature of this channel. In the denominator, when $k^{2}$ is small relative to $m_{N}\,\Delta^{(0)}$ an expansion of the dibaryon propagator is similar to the on-shell contact interactions in Ref.~\cite{Peng:2020nyz}. 

\begin{figure*}[t]
\centering
\includegraphics[scale=0.21]{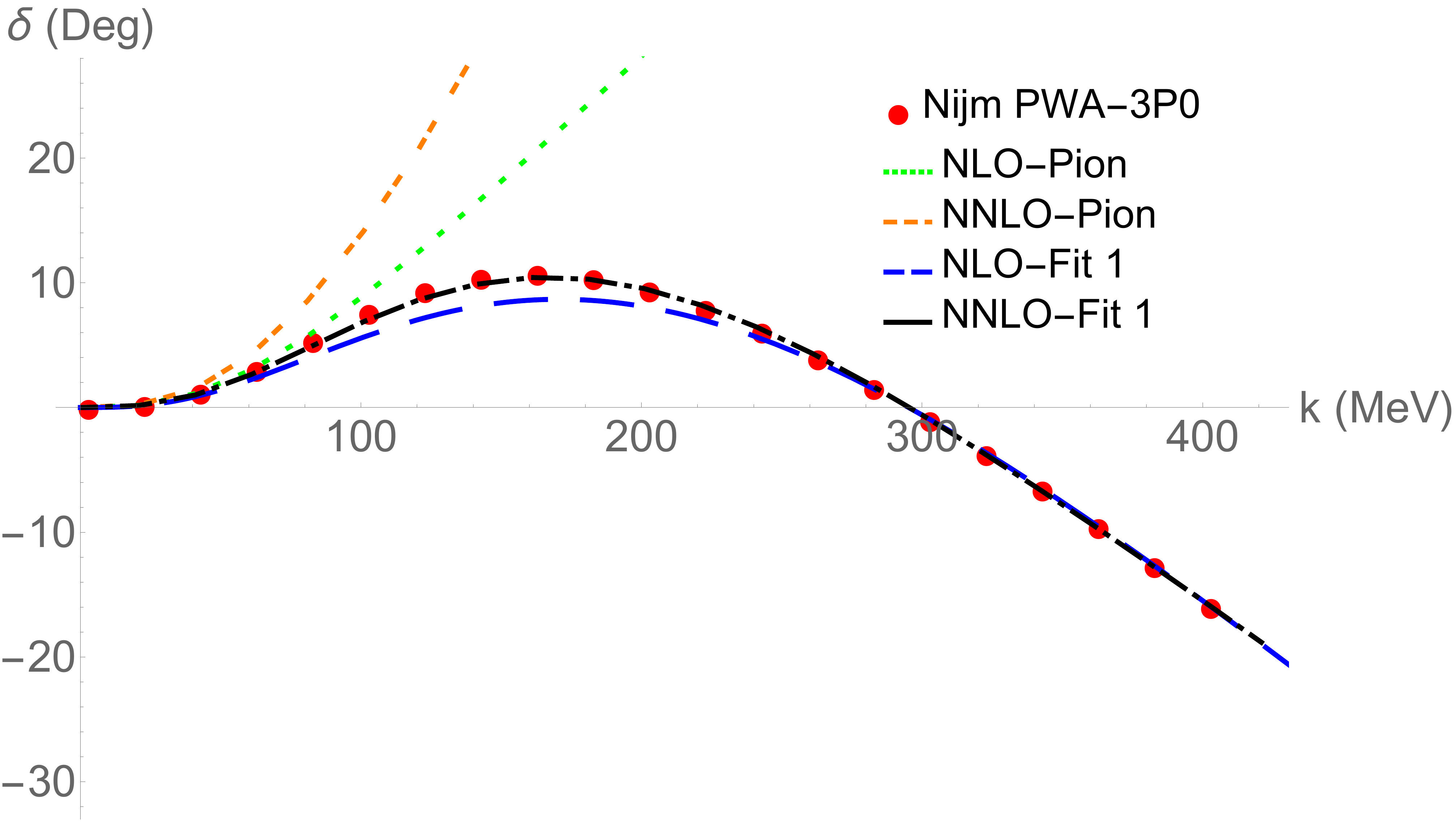}
\includegraphics[scale=0.21]{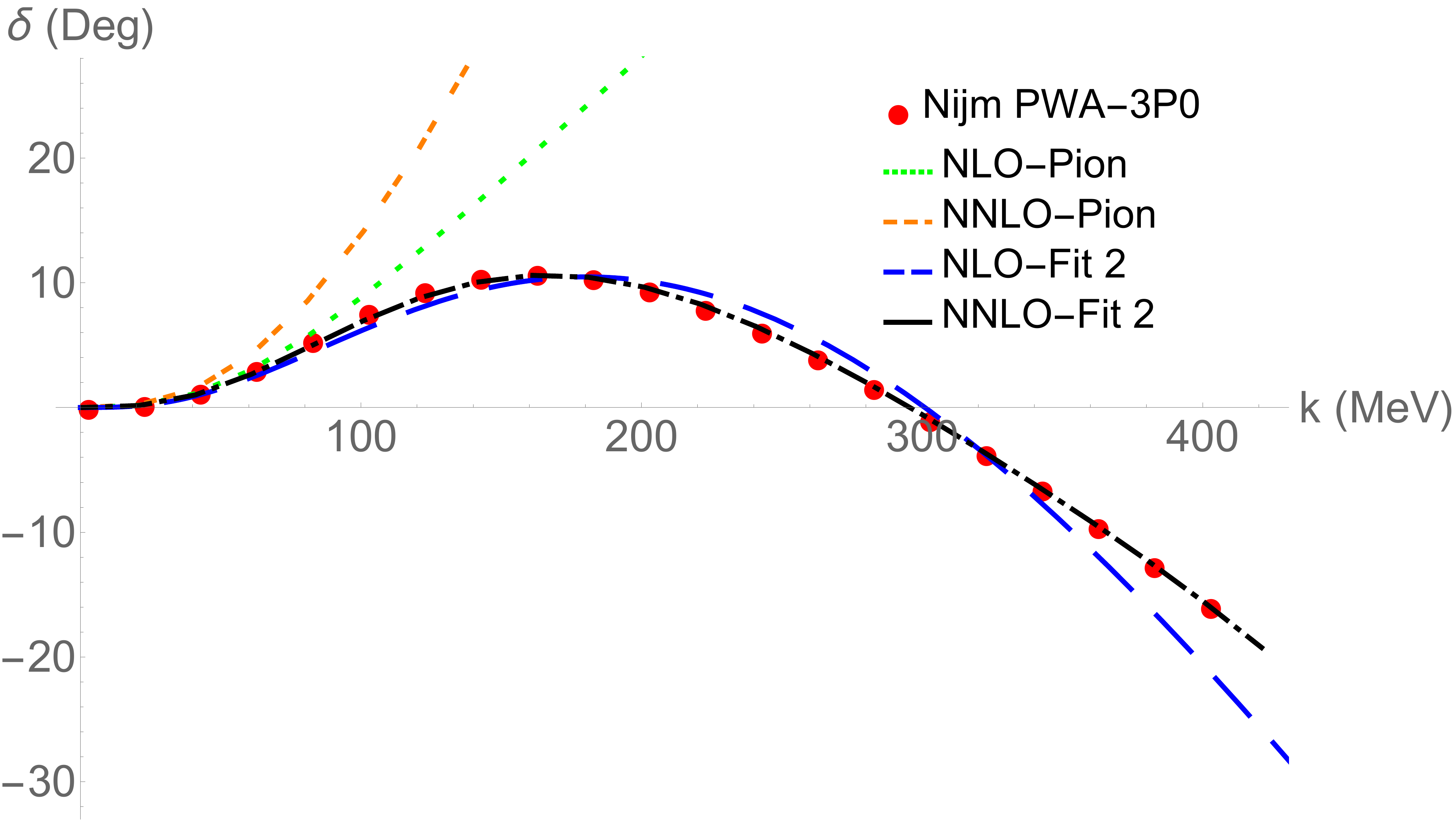}
\caption{The ${}^{3}\!P_{0}$ channel results for momentum sets of fit 1 (left) and fit 2 (right) in Table.~\ref{tbl.2}. For notation and explanation see \figref{fig.3}.}
\label{fig.4}
\end{figure*}

Fitted values of $g_{1}^{(1)}$, $\Delta^{(0)}$ and $C_{2}^{(2)}$ from two sets of momenta are given in Table.~\ref{tbl.2} and the phase shift is plotted in \figref{fig.4}. As we see from graphs, the importance of the dibaryon field shows itself in the $k^{2} \gg m_{N}\,\Delta^{(0)},\,m_{\pi}^2$ limit where the $T$ matrix is a constant at this order
\bea
\lim\limits_{k^{2} \gg m_{N}\,\Delta^{(0)},\,m_{\pi}^2}\! \overline{T}^{(1)\tiny(^{3}\!P_{0})}\!\!= - \frac{1}{\Lambda_{NN}} + \eta\,m_{N}\,{g_{1}^{(1)}}^2\!\!+ \ldots \, . \quad
\label{klimit3P0}
\eea
One can check that with $\eta = +1$ and fitted values of parameters in Table.~\ref{tbl.2}, sum of the first two terms is positive. Therefore, the phase shift decreases linearly at larger momenta, in contrast to the quadratic behavior due to the contact term $C_{2}\,k^2$ \cite{Peng:2020nyz}. 
\begin{table*}[t]
\centering
\begin{tabular*}{\textwidth}{@{\extracolsep{\fill} } c  c  c  c  c  c  c }
\hline\hline
${}^{3}\!P_{0}$  & $\{k_{1}, k_{2}, k_{3}\}$(MeV) & \; $g_{1}^{(1)}$ (MeV$^{-1}$)\; & \; $\Delta^{(0)}$ (MeV) \; & \; $C_{2}^{(2)}$ (MeV$^{-3}$) \; & \; $\sqrt{m_{N} |\Delta^{(0)}|}$ (MeV) \; & \; $\eta$ \;\\ \hline
Fit 1 & 300, 400, 200 & 0.00250 & -99.7 & 1.2$\times 10^{-8}$ & 305.9 & +1\\
Fit 2 & 180, 320, 380 & 0.00286 & -168.0 & 2.6$\times 10^{-8}$ & 397.2 & +1
\\\hline\hline
\end{tabular*}
\caption{
Results for fitting up-to-NNLO LECs in the ${}^{3}\!P_{0}$ channel. For notation and explanation see Table.~\ref{tbl.1}.} 
\label{tbl.2}
\end{table*}

Again, fitted values for this channel are close to the estimation of the PC. The NLO EFT phase shift shows real improvement relative to results of the KSW PC for momenta below $k \approx 200$ MeV, and the agreement with data improves at NNLO. Similar to the ${}^{1}\!P_{1}$ channel, there is a possible imaginary zero in the denominator of the $T$ matrix in the ${}^{3}\!P_{0}$ channel.

\begin{figure*}[t]
\centering
\includegraphics[scale=0.21]{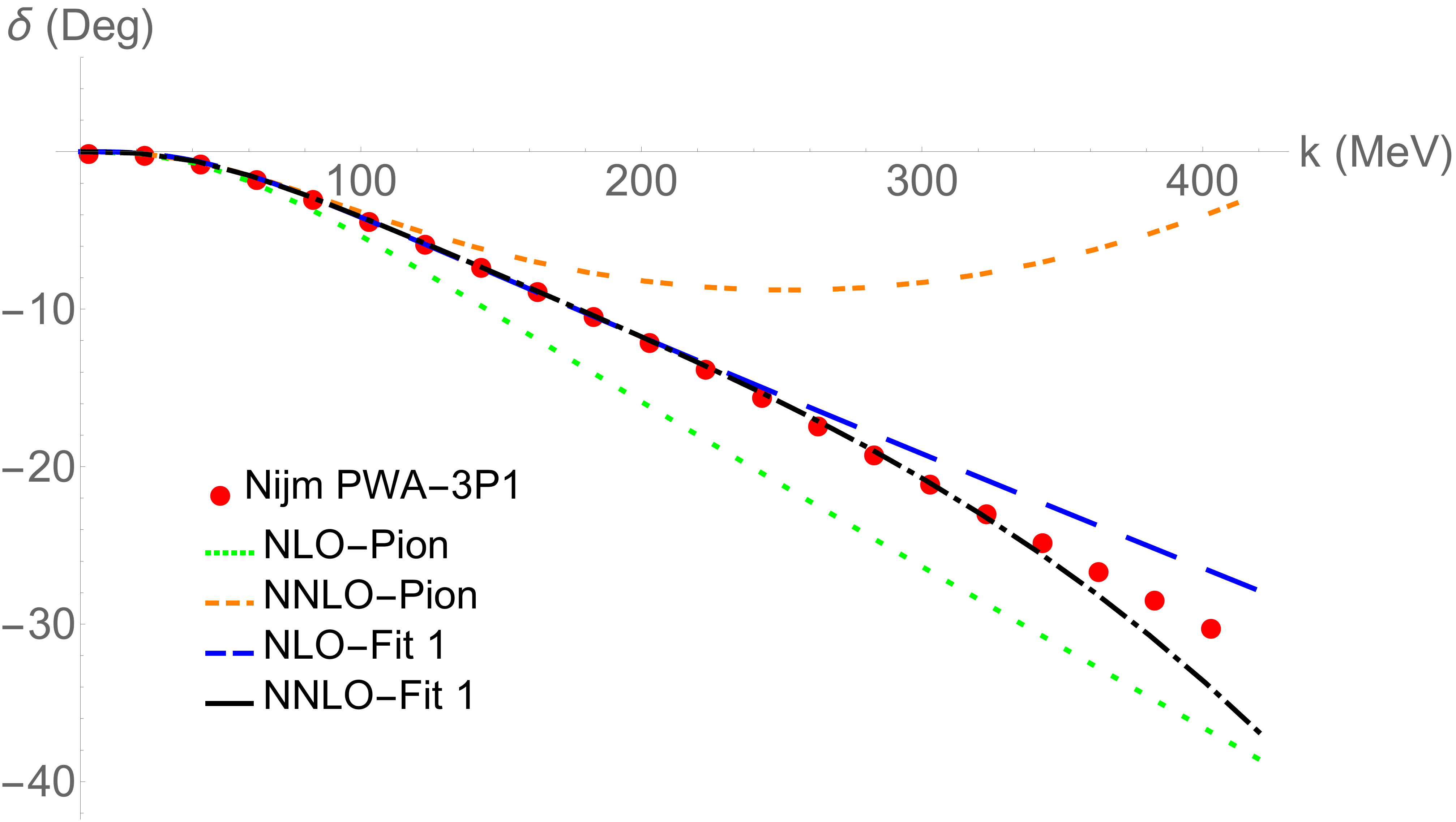}
\includegraphics[scale=0.21]{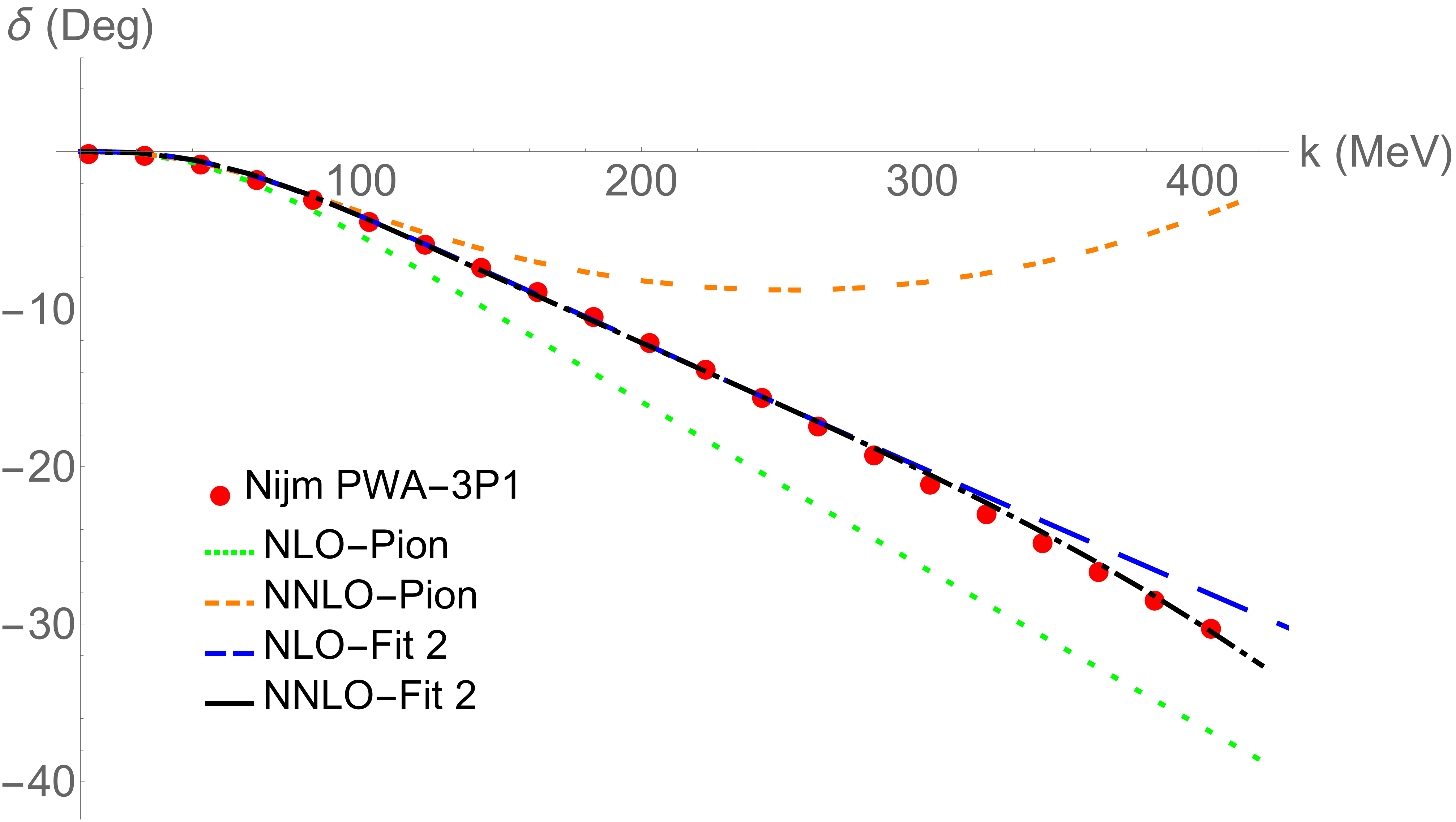}
\caption{The ${}^{3}\!P_{1}$ channel results for momentum sets of fit 1 (left) and fit 2 (right) in Table.~\ref{tbl.3}. For notation and explanation see \figref{fig.3}.}
\label{fig.5}
\end{figure*}

In the KSW PC there are no new parameters at NNLO in the $P$-wave channels, and therefore these channels are good benchmarks to test the convergence of the theory. For the ${}^{3}\!P_{1}$ channel, phase shifts at NLO and NNLO agree well with data only up to $k \approx 200$ MeV. This is evidence that pion interactions alone are not enough to get good agreement with data in a perturbative EFT for typical momenta about $300$ MeV. Alternative options are to either include additional contact interactions or a dibaryon field. In this paper, I consider the latter. The on-shell NLO $T$ matrix in the ${}^{3}\!P_{1}$ channel is (see Appendix~\ref{AppxA}) 
\bea
\overline{T}^{(1)\tiny({}^{3}\!P_{1})}\!\left(k, m_{\pi}^2\right)\!\! & = & \frac{\eta\,m_{N}\,{g_{1}^{(1)}}^2\,k^2}{k^{2}  - m_{N}\,\Delta^{(0)}}+\frac{1}{\Lambda_{NN}}\biggl[ \frac{2\,k^2 - m_{\pi}^2}{4\,k^2} + \frac{m_{\pi}^4}{16\,k^4}\,\ln\left(1 + \frac{4\,k^2}{m_{\pi}^2}\right) \biggr] \, .  \qquad
\eea
Fitted values of the LECs are given in Table.~\ref{tbl.3}, and the phase shift is plotted in \figref{fig.5}.  
\begin{table*}[t]
\centering
\begin{tabular*}{\textwidth}{@{\extracolsep{\fill} } c  c  c  c  c  c  c }
\hline\hline
${}^{3}\!P_{1}$ & $\{k_{1}, k_{2}, k_{3}\}$(MeV) & \; $g_{1}^{(1)}$ (MeV$^{-1}$)\; & \; $\Delta^{(0)}$ (MeV) \; & \; $C_{2}^{(2)}$ (MeV$^{-3}$) \; & \; $\sqrt{m_{N} |\Delta^{(0)}|}$ (MeV) \; & \; $\eta$ \;\\ \hline
Fit 1 & 60, 200, 300 & 0.00071 & -13.8 & 7.9$\times 10^{-9}$ & 114.0 & -1\\ 
Fit 2 & 100, 250, 390 & 0.00065 & -8.7 & 5.9$\times 10^{-9}$ & 90.4 & -1
\\\hline\hline
\end{tabular*}
\caption{
Results for fitting up-to-NNLO LECs in the ${}^{3}\!P_{1}$ channel. For notation and explanation see Table.~\ref{tbl.1}.} 
\label{tbl.3}
\end{table*}

Since $\Delta^{(0)} < 0$, there is a possible imaginary zero in this channel too, although unlike the other two channels $\eta = -1$. Again, size of the fitted parameters are of the same order as the estimation from the PC. In \figref{fig.5}, we see good agreement at NLO, which further improves at NNLO. Note that this happens because the dibaryon field cancels the large upturn due to the OPE box diagram in this channel. The large-$k$ limit of NLO $T$ matrix for this channel is
\bea
\lim\limits_{k^{2} \gg m_{N}\,\Delta^{(0)},\,m_{\pi}^2}\! \overline{T}^{(1)\tiny(^{3}\!P_{1})}\!\!= \frac{1}{2\,\Lambda_{NN}} + \eta\,m_{N}\,{g_{1}^{(1)}}^2\!\!+ \ldots \, , \quad
\label{klimit3P1}
\eea
and with the fitted parameters in Table.~\ref{tbl.2}, it is positive up to the first two terms. Again, this shows that phase shift decreases linearly for large momenta in this channel too.

\section{Conclusion}
\label{sec.5}
A new EFT including a dibaryon field with perturbative pions is investigated for the uncoupled $P$-wave channels of nucleon-nucleon scattering systems, and promising results up to NNLO are observed. This is (strong) evidence that to have a convergent perturbative EFT beyond the pion mass \cite{Kaplan:1999qa}, physics of higher energies has to be included by introducing an auxiliary dibaryon field to make pion interactions perturbative. The PC for chiral contact LECs in this theory are the same as NDA \cite{Manohar:1983md,Georgi:1992dw}, and the PC for CSB contact interactions in Eq.~\eqref{eq.1} has been adapted from the KSW PC and NDA. The PC for the chiral and CSB dibaryon LECs in Eqs.~\eqref{eq.2} and \eqref{eq.3} show that dibaryon interactions start perturbatively at NLO. 

Calculations are carried out up to NNLO for the selected $P$-wave channels with a new method that combines renormalization and fitting together. All pionless and related-to-dibaryon parts of diagrams are calculated analytically, although the finite part of the box diagram from iteration of the OPE potential is done numerically. No $m_{\pi}^2$ dependence has been observed in divergent parts, and therefore there is no need to add CSB ``contact" counter terms at NNLO. The CSB dibaryon LECs, however, start to appear at NNLO which absorb effect of nonanalytic terms proportional to $m_{\pi}^2$ coming from the iteration of the OPE potential. Since LECs at NLO and the new $C_{2}^{(2)}$ at NNLO have fixed values which do not run with the regulator, their sizes are estimated by the new PC in Eq.~\eqref{estimationLEC}. According to results given in Tables.~\ref{tbl.1}, \ref{tbl.2} and \ref{tbl.3}, fitted values of LECs are in good agreement with those estimated by PC. The renormalization method in Sec.~\ref{sec.3} is new and useful for numerical calculations, which uses data points instead of an interval of data. Furthermore, it produces renormalized parts that counteract effect of nonanalytic terms coming from pion interactions. In Appendix~\ref{AppxB}, the theory has been renormalized by using the usual method in the literature of nuclear EFTs, so one can compare results in both renormalization methods.

An advantage of including the dibaryon at NLO, instead of using contact interactions, is that the tree level $T$ matrix of the dibaryon is a constant in the large-$k$ limit ($k^{2} \gg m_{N}\,\Delta^{(0)},\,m_{\pi}^2$) as can be seen from Eqs.~\eqref{NLOphase}, \eqref{klimit1P1}, \eqref{klimit3P0} and \eqref{klimit3P1}. This means the phase shift at NLO has a linear behavior in this limit, which is also noticeable in the Nijmegen PWA of these channels for larger momenta below the mass of $\rho$ meson. 

These promising results for the uncoupled $P$-wave channels are encouraging and motivate the application of the same idea to the ${}^{1}\!S_{0}$, ${}^{3}\!S_{1}$-${}^{3}\!D_{1}$, ${}^{3}\!P_{2}$-${}^{3}\!F_{2}$ and higher partial wave channels. The $S$-wave channels are more challenging than others, because they have a nonzero LO $T$ matrix from the resummation of $C_{0}^{(0)}$.  Preliminary NLO calculations for higher partial wave channels also show promising results, although before carrying out NNLO calculations in those channels I cannot draw solid conclusions yet.

\begin{acknowledgments}
I would like to thank S.~Fleming and U.~van~Kolck for their encouragement and valuable support, useful discussions about renormalization and thoughtful comments on the manuscript. This research was supported in part by the U.S. Department of Energy, Office of Science, Office of Nuclear Physics, under Award No. DE-FG02-04ER41338.
\end{acknowledgments}
\appendix
\section{Ingredients for $T$ matrix calculations}
\label{AppxA}
One can drive off-shell tree level pion and dibaryon diagrams by using the trace technique explained in Ref.~\cite{Fleming:1999ee}. For the ${}^{1}\!P_{1}$, ${}^{3}\!P_{0}$ and ${}^{3}\!P_{1}$ channels, I get
\bea
\overline{T}_{\pi}^{(1)\tiny({}^{1}\!P_{1})}\!\left(p', p, k, m_{\pi}^2\right) &=& \frac{1}{\Lambda_{NN}}\Biggl[ -\frac{3\,m_{\pi}^2}{2 p \,p'} + \frac{3\,m_{\pi}^2\left(m_{\pi}^2 + p^2+p'^2\right)}{8\,p^2\,p'^2} \ln\left(\frac{m_{\pi}^2 + \left(p + p'\right)^2}{m_{\pi}^2 + \left(p - p'\right)^2}\right) \Biggr] \, , \qquad \\
\overline{T}_{\pi}^{(1)\tiny({}^{3}\!P_{0})}\!\left(p', p, k, m_{\pi}^2\right) &=& \frac{1}{\Lambda_{NN}}\Biggl[ -\frac{1}{2 p \,p'}\left(p^2+p'^2\right) + \frac{m_{\pi}^2\left(p^2+p'^2\right) + \left(p^2-p'^2\right)^2}{8\,p^2\,p'^2} \ln\left(\frac{m_{\pi}^2 + \left(p + p'\right)^2}{m_{\pi}^2 + \left(p - p'\right)^2}\right) \Biggr] \, , \qquad \\
\overline{T}_{\pi}^{(1)\tiny({}^{3}\!P_{1})}\!\left(p', p, k, m_{\pi}^2\right) &=& \frac{1}{\Lambda_{NN}}\Biggl[ \frac{1}{4 p \,p'}\left(p^2+p'^2 - m_{\pi}^2\right) + \frac{m_{\pi}^4 - \left(p^2-p'^2\right)^2}{16\,p^2\,p'^2}\ln\left(\frac{m_{\pi}^2 + \left(p + p'\right)^2}{m_{\pi}^2 + \left(p - p'\right)^2}\right) \Biggr] \, , \qquad \\
\overline{T}_{d}^{(1)(s)}\left(p', p, k\right) &=& \frac{\eta^{(s)}\,m_{N}\,{g_{1}^{(1)(s)}}^2\,p\,p'}{k^{2}  - m_{N}\,\Delta^{(0)(s)}} \; .
\eea
Structure of dibaryon parts of off-shell $T$ matrices are the same, but with different LECs. One can check that on-shell pion $T$-matrices are the same as those in Ref. \cite{Fleming:1999ee}. Also, the numerical calculation of the $T$ matrix for the box diagram of OPE potential by using above off-shell $T$ matrices and Eq.~\eqref{ProjectedLSE} agrees with results in Ref.~\cite{Kaiser:1997mw}.

After inserting various contributions of the off-shell $T$ matrix in the projected LSE in Eq.~\eqref{ProjectedLSE}, I use contour integration to find divergent and finite parts of loop integrals. If I define the dimensionless loop integral as $I_{\pi d} = I_{\pi d}^{[\mathtt{div}]} + I_{\pi d}^{[\mathtt{fin}]}$, I get
\bea
I_{\pi d}^{\tiny({}^{1}\!P_{1})}\!\! &=& \frac{3\,m_{\pi}^2}{2\,\Lambda_{NN}\,k^2}\left[ m_{\pi} - k\,\left(1+\frac{m_{\pi}^2}{2 k^2}\right) \tan^{-1}\left(\frac{2\,k}{m_{\pi}}\right)\right] + \frac{3\,i\,m_{\pi}^2}{2\,\Lambda_{NN}\,k}\left[ 1 - \left(\frac{1}{2}+\frac{m_{\pi}^2}{4 k^2}\right) \ln\left(1+\frac{4\,k^2}{m_{\pi}^2}\right)\right] \label{Idpi1P1}
\\ && \nn \\
I_{\pi d}^{\tiny({}^{3}\!P_{0})}\!\! &=& \frac{4\,L_{1}}{3\,\Lambda_{NN}} - \frac{m_{\pi}^2}{2\,\Lambda_{NN}\,k} \tan^{-1}\left(\frac{2\,k}{m_{\pi}}\right) + \frac{i\,k}{\Lambda_{NN}}\left[ 1 - \frac{m_{\pi}^2}{4 k^2} \ln\left(1+\frac{4\,k^2}{m_{\pi}^2}\right)\right] \label{Idpi3P0}
\\ && \nn \\
I_{\pi d}^{\tiny({}^{3}\!P_{1})}\!\! &=& -\frac{2\,L_{1}}{3\,\Lambda_{NN}} + \frac{m_{\pi}^3}{4\,\Lambda_{NN}\,k^2}\left[ 1 - \frac{m_{\pi}}{2\,k} \tan^{-1}\left(\frac{2\,k}{m_{\pi}}\right)\right] + \frac{i}{2\,\Lambda_{NN}\,k}\!\left[ \frac{m_{\pi}^2}{2} - k^2 - \frac{m_{\pi}^4}{8 k^2}\ln\left(1+\frac{4\,k^2}{m_{\pi}^2}\right)\right] \label{Idpi3P1} \qquad
\eea
where $L_{m}$ are given in Eq.~\eqref{Lm}. It is interesting that in the ${}^{1}\!P_{1}$ channel the cross pion-dibaryon loop integral does not contain a divergent term, although from a naive counting one expects the opposite. The reason is the cancellation among integrand terms in the on-shell NNLO $T$ matrix. As we can see in above equations, real and finite parts of pion-dibaryon loop ingetrals are complicated functions of $k/m_{\pi}$ and also contain odd powers of $m_{\pi}$ and negative powers of $k$. There are no terms in the Lagrangian with odd powers of $m_{\pi}$ or negative powers of $k$, and therefore I absorb effects of these nonanalytic terms at specific momenta into the same order LECs in the renormalization step (see Sec.~\ref{sec.3}). Another approach is to absorb effects of these terms into a redefinition of NLO LECs during fitting (see Appendix~\ref{AppxB}).

Running of NNLO contributions of NLO LECs in Sec.~\ref{sec.3} with regulator $\Lambda$ is given by
\bea
\tilde{\Delta}^{(1)} &=& \Delta^{(1)} + \omega_{2}^{(1)}\,m_{\pi}^2 \nn \\
&=& \eta\,{g_{1}^{(1)}}^2 \left( L_{3} + m_{N}\,\Delta^{(0)}\,L_{1}\right) + \theta^{(1)}\,m_{\pi}^2 + \frac{\left(k_{1}^2 - m_{N}\,\Delta^{(0)}\right)\left(k_{2}^2 - m_{N}\,\Delta^{(0)}\right)}{{g_{1}^{(1)}}^2\,m_{N}^2}\,\eta\,C_{2}^{(2)} \nn \\
&& +\ \frac{\left(k_{1}^2 - m_{N}\,\Delta^{(0)}\right)^2\left(k_{2}^2 - m_{N}\,\Delta^{(0)}\right)}{{g_{1}^{(1)}}^2\,m_{N}^2\,k_{1}^2\,\left(k_{1}^2 - k_{2}^2\right)}\,\eta\,\mathcal{R}^{(2)}_{k_{1}} - \frac{\left(k_{1}^2 - m_{N}\,\Delta^{(0)}\right)\left(k_{2}^2 - m_{N}\,\Delta^{(0)}\right)^2}{{g_{1}^{(1)}}^2\,m_{N}^2\,k_{2}^2\,\left(k_{1}^2 - k_{2}^2\right)}\,\eta\,\mathcal{R}^{(2)}_{k_{2}}\, ,\label{eq.A8}\qquad
\\ && \nn \\
\tilde{g}_{1}^{(2)} &=& g_{1}^{(2)} + h_{3}^{(2)}\,m_{\pi}^2 \nn \\
&=& \frac{\eta}{2}\,{g_{1}^{(1)}}^3\,m_{N}\,L_{1} - g_{1}^{(1)}\,I_{\pi d}^{[\mathtt{div}]} + \frac{\gamma^{(2)}\,m_{\pi}^2}{2\,g_{1}^{(1)}} + \frac{2 m_{N}\,\Delta^{(0)} - \left(k_{1}^2 + k_{2}^2\right)}{2 g_{1}^{(1)}\,m_{N}}\,\eta\,C_{2}^{(2)}	 \nn \\
&& -\ \frac{\left(k_{1}^2 - m_{N}\,\Delta^{(0)}\right)^2}{2 g_{1}^{(1)}\,m_{N}\,k_{1}^2\,\left(k_{1}^2 - k_{2}^2\right)}\,\eta\,\mathcal{R}^{(2)}_{k_{1}} + \frac{\left(k_{2}^2 - m_{N}\,\Delta^{(0)}\right)^2}{2 g_{1}^{(1)}\,m_{N}\,k_{2}^2\,\left(k_{1}^2 - k_{2}^2\right)}\,\eta\,\mathcal{R}^{(2)}_{k_{2}} \, , \label{eq.A9}
\eea
where values of $\Delta^{(0)}$, $g_{1}^{(1)}$ and $C_{2}^{(2)}$ are the fitted values in the tables. The $\mathcal{R}^{(2)}_{k_{1,2}} \equiv \mathcal{R}^{(2)}\left(k_{1,2}, g_{1}^{(1)}, \Delta^{(0)}, m_{\pi}^2\right)$ functions contain chiral and CSB parts, and therefore all the chiral terms of right-hand sides in above equations will define $\Delta^{(1)}$ and $g_{1}^{(2)}$. Also, all CSB parts will define $\omega_{2}^{(1)}\,m_{\pi}^2$ and $h_{3}^{(2)}\,m_{\pi}^2$. These chiral and CSB LECs absorb nonanalytic parts coming from $\mathcal{R}^{(2)}_{k_{1,2}}$ \cite{Kaplan:1999qa,Rupak:1999aa}. 
\begin{table*}[t]
\centering
\begin{tabular*}{\textwidth}{@{\extracolsep{\fill} } c  c  c  c  c  c  c  c  c  c  c  c }
\hline\hline
${}^{1}\!P_{1}$ & $\{k_{1}, k_{2}, k_{3}\}$ & $g_{1}^{(1)}$ & $\bar{g}_{1}^{(1)}$ & $\%\,g_{1}^{(1)}$ & $\Delta^{(0)}$ & $\bar{\Delta}^{(0)}$ & $\%\Delta^{(0)}$ & $C_{2}^{(2)}$ & $\sqrt{m_{N} |\Delta^{{}^{(0)}}|}$ & $\sqrt{m_{N} |\bar{\Delta}^{{}^{(0)}}|}$ &  $\eta$ \\ \hline
Fit 1 & 350, 400, 300 & 0.00112 & 0.00195 & 74.6 & -97.8 & -153.2 &  56.7 & -4.3$\times 10^{-9}$ & 303.0 & 379.2 & +1\\
Fit 2 & 310, 370, 280 & 0.00123 & 0.00292 &  137.8 & -149.2 & -304.9 & 104.3 & -1.1$\times 10^{-8}$ & 374.3 & 535.0 & +1 \\ \hline\hline
\end{tabular*}
\caption{Results for fitting up to NNLO LECs of the ${}^{1}\!P_{1}$ channel with the alternative method of renormalization. I refit total LECs given in Eqs.~\eqref{deltabar}-\eqref{g1bar} with the same momentum sets. The $\%\,g_{1}^{(1)}$ and $\%\Delta^{(0)}$ show the percentage of change relative to NLO values after re-fitting. The units are the same as Table.~\ref{tbl.1}.} 
\label{tbl.4}
\end{table*}

\section{The Second Method of Renormalization}
\label{AppxB}
\begin{table*}[t]
\centering
\begin{tabular*}{\textwidth}{@{\extracolsep{\fill} } c  c  c  c  c  c  c  c  c  c  c  c }
\hline\hline
${}^{3}\!P_{0}$ & $\{k_{1}, k_{2}, k_{3}\}$ & $g_{1}^{(1)}$ & $\bar{g}_{1}^{(1)}$ & $\%\,g_{1}^{(1)}$ & $\Delta^{(0)}$ & $\bar{\Delta}^{(0)}$ & $\%\Delta^{(0)}$ & $C_{2}^{(2)}$ & $\sqrt{m_{N} |\Delta^{{}^{(0)}}|}$ & $\sqrt{m_{N} |\bar{\Delta}^{{}^{(0)}}|}$ &  $\eta$ \\ \hline
Fit 1 & 300, 400, 200 & 0.00250 & 0.00359 & 44.0 & -99.7 & -75.6 &  24.2 & 1.0$\times 10^{-8}$ & 305.9 & 266.4 & +1\\ 
Fit 2 & 180, 320, 380 & 0.00286 & 0.00373 &  29.6 & -168.0 & -85.5 &  49.1 & 8.5$\times 10^{-9}$ & 397.2 & 283.3 & +1
\\\hline\hline
\end{tabular*}
\caption{
Results for fitting up to NNLO LECs of the ${}^{3}\!P_{0}$ channel with the alternative method of doing renormalization. For units, notation and explanation see Table.~\ref{tbl.4}.} 
\label{tbl.5}
\end{table*}
I can renormalize by not absorbing nonanalytic functions of $m_{\pi}$ into the NNLO LECs \cite{Kaplan:1998tg,Kaplan:1998we,Fleming:1999ee,Mehen:1998zz}. By setting square bracket terms in Eq.~\eqref{TNNLO1} equal to the same form as the right-hand side of Eq.~\eqref{2conditions} for a general $k$ instead of $k_{1,2}$, the NNLO $T$ matrix is
\bea
\overline{T}^{(2)} \left(k, m_{\pi}^2\right) &=& - i\,k\;\left[\overline{T}^{(1)}\left(k, m_{\pi}^2\right)\right]^2 + \frac{\eta\,m_{N}\,\gamma^{(2)}\,m_{\pi}^2\,k^2}{k^{2} - m_{N}\,\Delta^{(0)}} + \frac{\eta\,m_{N}^2\,{g_{1}^{(1)}}^2\,\theta^{(1)}\,m_{\pi}^2\,k^2}{\left(k^{2}  - m_{N}\,\Delta^{(0)}\right)^2} + C_{2}^{(2)}\,k^2 \nn\\ 
&& +\ \mathcal{R}^{(2)}\left(k, g_{1}^{(1)}, \Delta^{(0)}, m_{\pi}^2\right)
\, .\label{finalNNLOTR2}
\eea
Bare NNLO LECs are given by Eqs.~\eqref{eq.A8} and \eqref{eq.A9}, after removing terms containing $C_{2}^{(2)}$ and $\mathcal{R}^{(2)}_{k_{1,2}}$. The total up to NNLO phase shift also has a new form,
\bea
-\frac{\delta^{(t)}}{k} &=& -\frac{1}{k}\!\left(\!\delta^{(1)} + \delta^{(2)}\!\right) = \overline{T}^{(1)}_{\pi}\left(k, m_{\pi}^2\right) + \frac{\eta  \,m_{N}\,\bar{g}{_{1}^{(1)}}^2\,k^2}{k^{2}  - m_{N}\,\bar{\Delta}^{(0)}} + C_{2}^{(2)}\,k^2 + \mathcal{R}^{(2)}\!\left(k, g_{1}^{(1)}, \Delta^{(0)}, m_{\pi}^2\right) , \qquad
\label{totalPhaseR2}
\eea
where the $\bar{\Delta}^{(0)}$ and $\bar{g}_{1}^{(1)\,2}$ are given in Eqs.~\eqref{deltabar} and \eqref{g1bar}. The chiral part of NNLO $T$ matrix does not vanish at $k = k_{1}, k_{2}$, and therefore unlike the previous method of renormalization I let $\bar{\Delta}^{(0)}$ and $\bar{g}_{1}^{(1)\,2}$ change when fitting at NNLO. I use NLO fitted values of $\Delta^{(0)}$ and $g_{1}^{(1)\,2}$ in the NNLO part of the phase shift, and $\bar{\Delta}^{(0)}$ and $\bar{g}_{1}^{(1)\,2}$ are used in the NLO part only. Since there are nonanalytic terms I cannot predict effects of these terms on refitted LECs because $Q \sim k \sim m_{\pi} \sim \Lambda_{NN}$. These effects can be small or large relative to NLO values, but the only thing that I expect is that even refitted values are within the estimation of the PC in Eq.~\eqref{estimationLEC}. 
\begin{figure*}[b]
\centering
\includegraphics[scale=0.21]{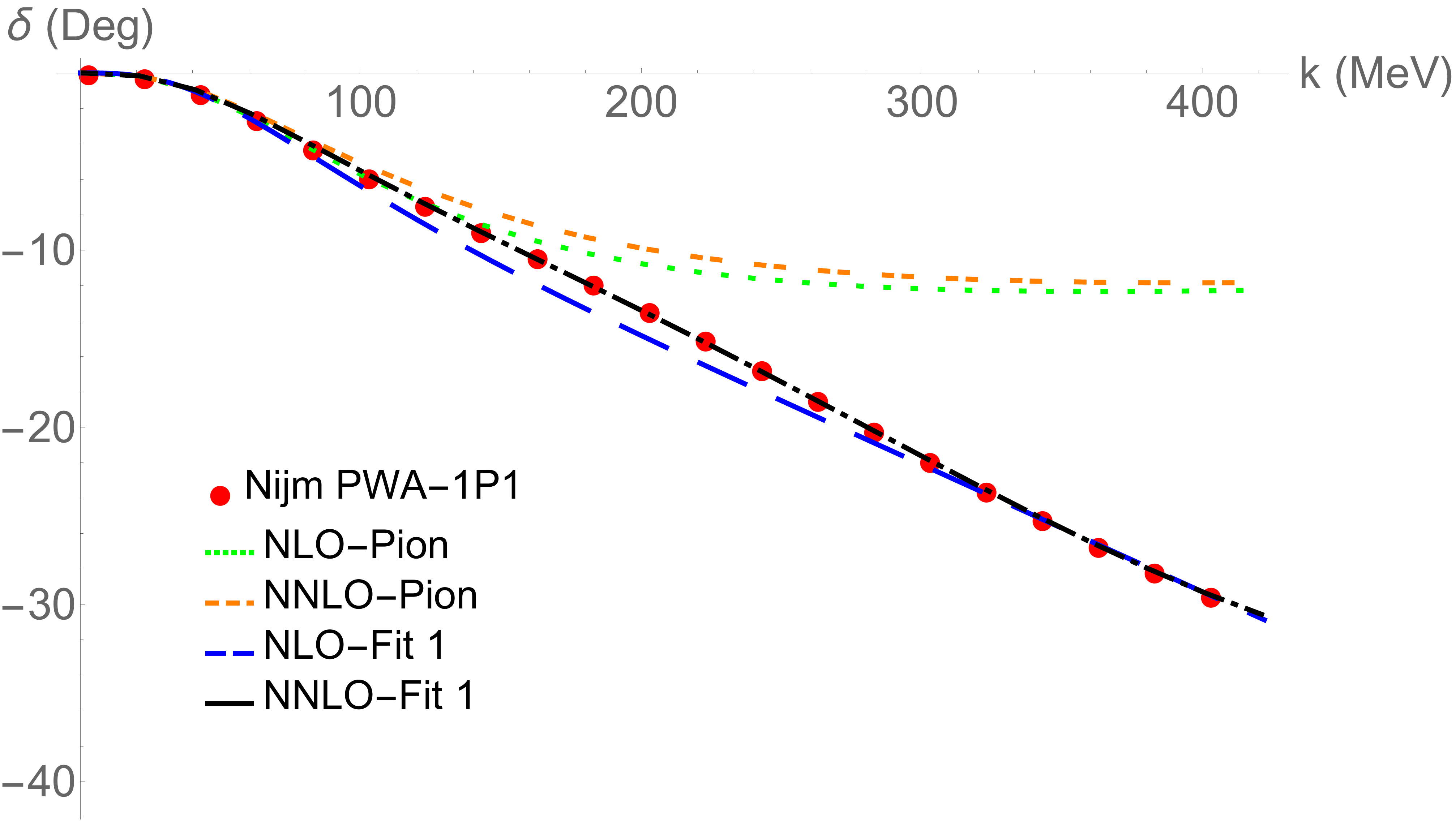}
\includegraphics[scale=0.21]{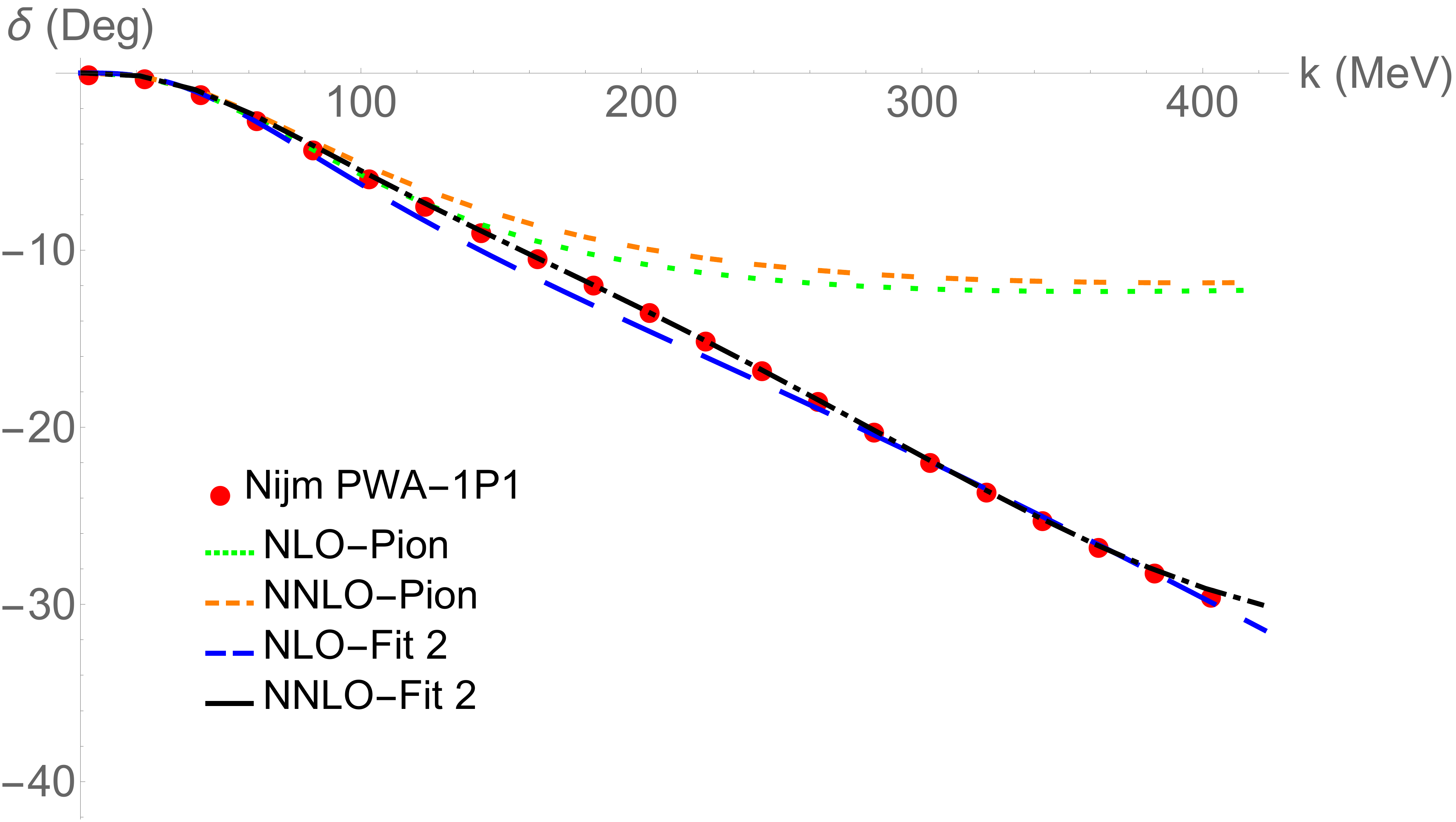}
\caption{The ${}^{1}\!P_{1}$ channel results from an alternative method of renormalization for momentum sets of fit 1 (left) and fit 2 (right) in Table.~\ref{tbl.4}. For notation and explanation see \figref{fig.3}.}
\label{fig.6}
\end{figure*}

\begin{figure*}[t]
\centering
\includegraphics[scale=0.21]{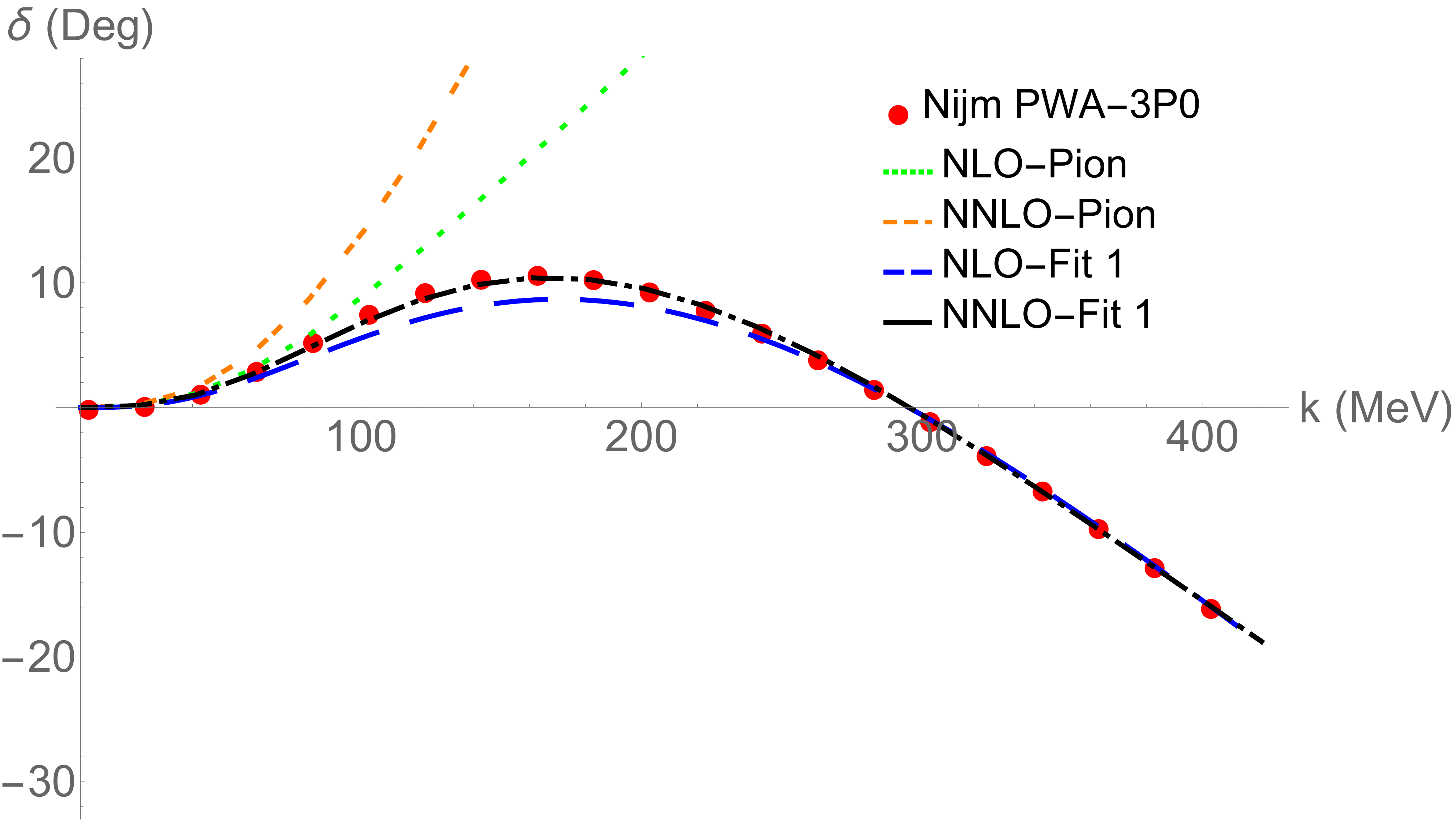}
\includegraphics[scale=0.21]{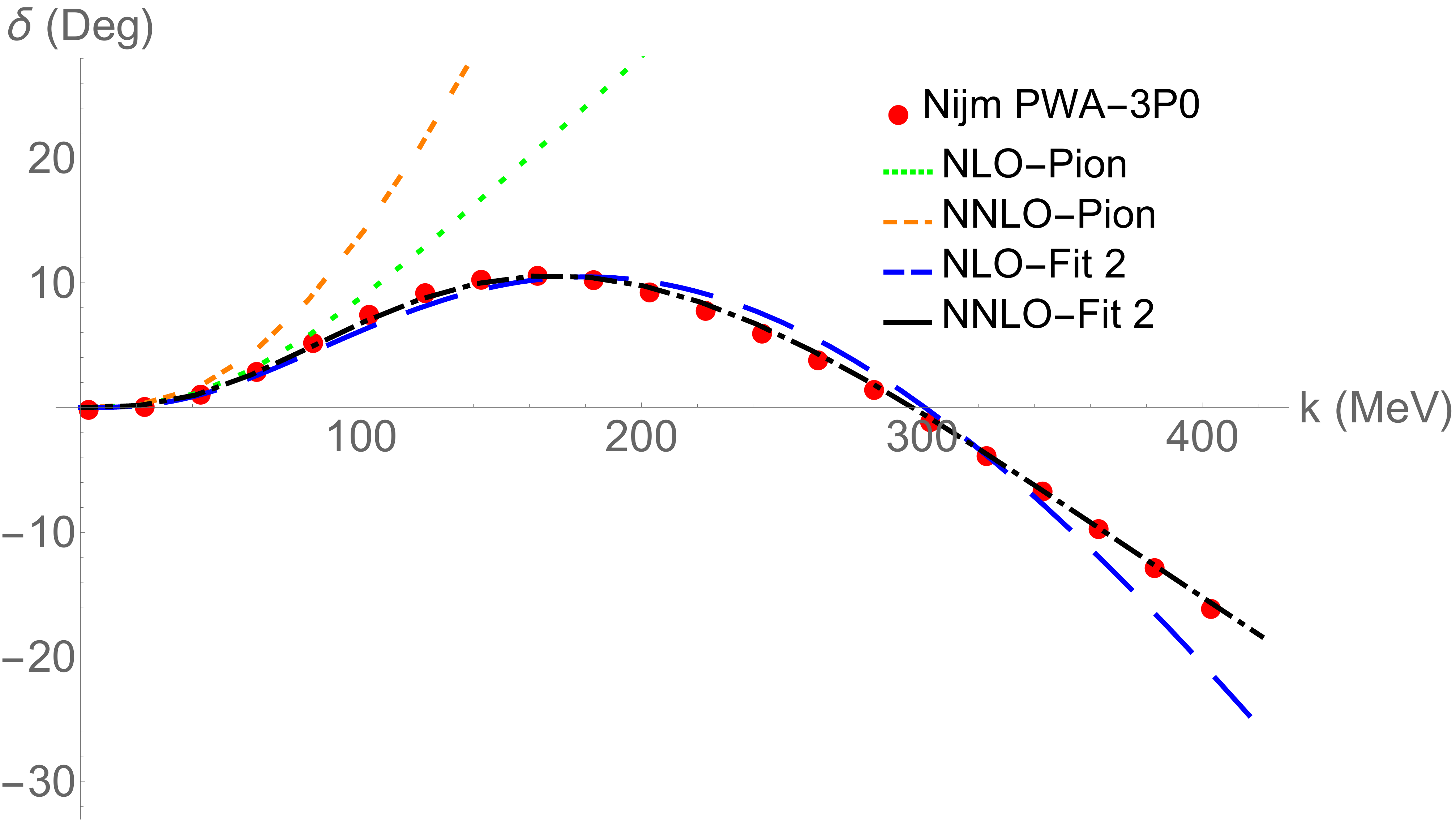}
\caption{The ${}^{3}\!P_{0}$ channel results from an alternative method of renormalization for momentum sets of fit 1 (left) and fit 2 (right) in Table.~\ref{tbl.5}. For notation and explanation see \figref{fig.3}.}
\label{fig.7}
\end{figure*}

Results for ${}^{1}\!P_{1}$ and ${}^{3}\!P_{0}$ channels are given in Tables.~\ref{tbl.4} and \ref{tbl.5} and plots of the phase shift are shown in Figs.~\ref{fig.6} and \ref{fig.7}, respectively. In the ${}^{3}\!P_{1}$ channel, the error introduced by using data points is too high, which makes it hard to find reasonable refitted values. For this case, it is better to use an interval of data and for that an analytical expression of $\mathcal{R}^{(2)}\left(k, g_{1}^{(1)}, \Delta^{(0)}, m_{\pi}^2\right)$ is preferable.

As we can see in Tables.~\ref{tbl.4} and \ref{tbl.5}, effects of nonanalytic terms that have been absorbed in refitted values are not perturbative and small relative to NLO fitted values, although generally results at NNLO are in better agreement with data than at NLO. As I expected, refitted values are also within the estimation of PC in Eq.~\eqref{estimationLEC}.

Note that my goal is not to find a PC in which contributions from the OPE potential are small, and therefore it is not appropriate to conclude from the large shift in fitted values of LECs going from NLO to NNLO that perturbation theory breaks down. My goal is to make effects of pion interactions perturbative by using the dibaryon field and indeed this is happening in the first method of renormalization. In the first method, however, I cannot find the change to fitted NLO parameters until I know values of $\gamma^{(2)}$ and $\theta^{(1)}$.


\end{document}